\documentclass[11pt]{article}
\usepackage{amsmath,amssymb}
\usepackage{graphicx}
\usepackage{subcaption}
\usepackage{xcolor}
\usepackage{caption}

\begin{document}

\setcounter{page}{1}

\pagestyle{plain}

\begin{center}
\Large{\bf Anisotropic Dirac-Born-Infeld Inflation with Non-Vacuum Initial States: Primordial Perturbations, Non-Gaussianity, and Observational Constraints}\\
\small \vspace{1cm} {\bf Narges Rashidi
$^{}$\footnote{n.rashidi@umz.ac.ir (Corresponding Author) }} and {\bf Maryam Roushan
$^{}$\footnote{m.roushan@umz.ac.ir }}   \\
\vspace{0.5cm} Department of Theoretical Physics, Faculty of
Science,
University of Mazandaran,\\
P. O. Box 47416-95447, Babolsar, IRAN\\
\end{center}

\begin{abstract}
	
We investigate linear and nonlinear primordial perturbations in an
anisotropic Dirac-Born-Infeld (DBI) inflationary model with a
non-vacuum initial state. Using the Arnowitt-Deser- Misner (ADM) formalism, we expand the
action up to second and third order in the curvature perturbation
and derive the corresponding scalar and tensor power spectra, as
well as the bispectrum and the equilateral non-linearity parameter
\(f_{NL}^{\mathrm{equil}}\). The effects of anisotropic
corrections and non-Bunch-Davies (non-BD) initial conditions are incorporated
through the slow-roll sector and Bogoliubov coefficients. For the
numerical analysis, we consider an intermediate expansion scenario
together with a phenomenological ansatz for the excited-state
occupation number \(N_k\). By comparing the model predictions with
recent observational datasets, including Planck2018 TT, TE, EE +
lowE + lensing + BK18 + BAO and DESI+CMB+DESY5 data, we identify
observationally viable regions in the parameter space of the model.
Our analysis indicates that the anisotropic DBI scenario with
non-vacuum initial conditions can remain compatible with current
constraints on the scalar spectral index, tensor-to-scalar ratio,
and equilateral non-Gaussianity for suitable ranges of the
anisotropy parameter \(c\) and the initial-state parameter
\(N_{k,0}\).
\end{abstract}
\newpage

\section{\label{s1}Introduction}

The theory of inflation has become a foundational element of modern
cosmology, providing a plausible explanation for the large-scale
homogeneity and near-flat geometry observed in the universe. First
introduced in~\cite{Gut81}, inflation describes an early period of
accelerated expansion, typically driven by a scalar field known as the
inflaton~\cite{Lin82,Alb82,Lin90,Lid00a,Lid97}. This mechanism naturally
addresses several limitations of the classical Big Bang model, such as
the horizon and flatness problems, and simultaneously provides a
framework for the generation of primordial fluctuations~\cite{Muk81}.
These fluctuations later evolve into the cosmic structures observed
today. In its minimal slow-roll single-field realization, inflation
predicts curvature perturbations that are approximately scale invariant
and nearly Gaussian~\cite{Mal03}. Observations of the Cosmic Microwave Background (CMB) and Large-Scale Structure (LSS)
strongly support this paradigm, reinforcing its status as a cornerstone
of early-universe physics.

Over time, various inflationary models have been formulated to
account for the early time expansion of the universe, differing
primarily in the nature of the inflaton potential and the underlying
dynamics. Among these are chaotic inflation~\cite{Lin83}, hybrid
inflation~\cite{Lin94}, and k-inflation~\cite{Arm99}, each offering
a distinct theoretical route to realize inflation. A notable example
that emerges from string theory considerations is the
Dirac-Born-Infeld (DBI) inflation model~\cite{Sil04,Ali04}, in which
the inflaton corresponds to the location of a D-brane traversing a
compactified geometry with nontrivial warping. The DBI framework
modifies the inflaton's kinetic term by introducing a relativistic
correction, effectively limiting the speed of its evolution. This
unique feature leads to characteristic observational signatures,
including potentially large non-Gaussianities in the primordial
fluctuations, which make the DBI model a well-motivated scenario for
exploring early-universe physics.

Although most conventional inflationary scenarios are formulated in
an isotropic background, models with small deviations from exact
isotropy have also attracted considerable attention. Such scenarios
provide a broader framework for investigating possible departures
from the standard inflationary picture and their implications for
primordial perturbations. A common approach to generating
anisotropic effects during inflation is to couple the inflaton field
to additional vector or gauge degrees of freedom that can mildly
break rotational invariance. For example, a kinetic coupling of the
form $(f(\varphi))^2F_{\mu\nu}F^{\mu\nu}$ allows gauge fields to
remain dynamically relevant throughout inflation due to the time
dependence of the coupling function
$f(\varphi)$~\cite{Bar13}. Similar mechanisms have been studied in
several contexts, including anisotropic inflation in supergravity
models~\cite{Kan10}, models involving vector and two-form
fields~\cite{Oh13}, and inflationary scenarios with Gauss-Bonnet
corrections~\cite{La16}. Anisotropic inflation has also been investigated in several non-canonical frameworks, including anisotropic power-law inflation in DBI theories~\cite{Do21a}, anisotropic power-law k-inflation~\cite{oh13} and anisotropic inflation with derivative couplings~\cite{Ho18}. Additional extensions include constant-roll
anisotropic inflation~\cite{Ng21,It18} and multi-field models containing
both scalar and vector degrees of freedom with detailed stability
analyses~\cite{Do21}.

Such frameworks can introduce small anisotropic corrections to the
inflationary dynamics and consequently modify the properties of
primordial perturbations. In particular, they may affect the scalar
and tensor observables as well as the amplitude and shape of
non-Gaussian signals. Although current observations strongly
constrain deviations from exact isotropy, anisotropic inflationary
models remain useful laboratories for studying possible
symmetry-breaking effects in the early universe. Exploring such
extensions broadens the landscape of viable inflationary scenarios
and provides additional ways to test early-universe physics through
precision cosmological observations.

In addition to anisotropic effects, deviations from the standard Bunch-Davies vacuum have also attracted considerable attention, since excited initial states can modify both the primordial power spectra and the resulting non-Gaussian signatures~\cite{Che07} (see also \cite{Hol08}). In this context, it is interesting to investigate how anisotropic background corrections and non-vacuum initial conditions influence both linear and nonlinear primordial perturbations within non-canonical inflationary frameworks.
In the present work, we consider an anisotropic DBI
inflationary scenario in the presence of a non-vacuum initial state,
focusing on the effects of anisotropy and excited initial conditions
on the scalar and tensor power spectra as well as on the equilateral
non-Gaussianity parameter. Using recent observational datasets, we
also examine the observational viability of the model and constrain
the allowed regions of the anisotropy and non-vacuum parameters.

This study builds upon the framework proposed
in~\cite{Noj22}, where anisotropic cosmological backgrounds were investigated. In the present work, we employ the Bianchi type I metric given in Eq.~(2) and study a DBI inflationary scenario in the presence of anisotropic background corrections and a non-vacuum initial state. In Section~\ref{s2}, we derive the background field
equations and the corresponding slow-roll parameters in terms of the
anisotropic variables. In Section~\ref{s3}, we study scalar and
tensor perturbations and obtain the modified power spectra in the
presence of anisotropic contributions. Section~\ref{s4} is devoted
to the analysis of primordial non-Gaussianity and the resulting
equilateral non-linearity parameter. In
Section~\ref{s5}, we perform a numerical and observational analysis
using recent cosmological datasets~\cite{Pl18a,Pl19,Pa22,Wa24,Co24}
to constrain the model parameters and examine the observational
viability of the scenario. Finally, Section~\ref{s6} summarizes the
main results and presents the concluding remarks.

\section{\label{s2}The Background Setup of the Model}
We begin our analysis with the DBI inflationary model given by the following action
\begin{eqnarray}
    \label{eq1} S=\int d^{4}x\,\sqrt{-g}
    \left[\frac{\cal{R}}{2\kappa^{2}}-{\cal{T}}^{-1}(\varphi)\sqrt{1+{\cal{T}}(\varphi)\,\partial_{\mu}\varphi\,\partial^{\mu}\varphi}-V(\varphi)\right]\,,
\end{eqnarray}
where $\mathcal{R}$ is the Ricci scalar, $\kappa$ denotes the
gravitational constant, ${\cal{T}}^{-1}(\varphi)$ characterizes the
effective brane tension, and $V(\varphi)$ is the potential term
corresponding to the DBI field. We note that Eq. (1) follows the convention adopted in the original DBI inflation formulation of Alishahiha, Silverstein and Tong \cite{Ali04}, in which the DBI scalar-field sector is written without an explicit subtraction term proportional to ${\cal T}^{-1}(\varphi)$ separately. Similar forms have also been employed in subsequent studies of DBI inflation \cite{To04,Li13}. Throughout this work, we adopt this convention consistently.  

We consider a homogeneous but anisotropic Bianchi type I spacetime background given by~\cite{Noj22}
\begin{eqnarray}
    \label{eq2}
    ds^2=-dt^2+a(t)^{2}\,\sum_{j=1}^{3}\,\exp(2\zeta_{j}(t))\,(dx^{j})^{2}\,,
\end{eqnarray}
in which $\zeta_j(t)$ parameterizes the anisotropy in each spatial
direction. As discussed in~\cite{Noj22}, we define the average of
anisotropy parameters as
\begin{eqnarray}
    \label{eqn3}
    \bar{\zeta}(t) = \frac{1}{3} \sum_{j=1}^3 \zeta_j(t),\nonumber
\end{eqnarray}
and perform the redefinitions
\begin{eqnarray}
    \label{eqn4}
    a(t) \rightarrow a(t)\, e^{\bar{\zeta}(t)}, \quad \zeta_j(t) \rightarrow \zeta_j(t) - \bar{\zeta}(t),\nonumber
\end{eqnarray}
which lead to the following constraints
\begin{eqnarray}
    \label{eqn5}
    \sum_{j=1}^3 \zeta_j(t) = 0, \qquad \sum_{j=1}^3 \dot{\zeta}_j(t) = 0.\nonumber
\end{eqnarray}

In this framework, the modified Friedmann equations take the form
\begin{eqnarray}
    \label{eq3}
    H^2=\frac{\kappa^{2}}{3}\left[\frac{{\cal{T}}^{-1}}{\sqrt{1-{\cal{T}}\dot{\varphi}^{2}}}+V(\varphi)\right]+\frac{1}{6}\sum_{i=1}^{3}(\dot{\zeta}^{i})^{2}\,,
\end{eqnarray}
\begin{eqnarray}
    \label{eq4}
    2\dot{H}+3H^{2}=-\kappa^{2}\,\left[-{\cal{T}}^{-1}\,\sqrt{1-{\cal{T}}\dot{\varphi}^{2}}-V(\varphi)\right]-\frac{1}{2}\sum_{i=1}^{3}(\dot{\zeta}^{i})^2\,,
\end{eqnarray}
The scalar field $\varphi$ evolves according to
\begin{eqnarray}
    \label{eq5}
    \frac{\ddot{\varphi}}{(1-{\cal{T}}\dot{\varphi}^{2})^{\frac{3}{2}}}+\frac{3H\dot{\varphi}}{(1-{\cal{T}}\dot{\varphi}^{2})^{\frac{1}{2}}}+V'=
    \frac{{\cal{T}}'}{{\cal{T}}^{2}}\frac{2-3{\cal{T}}\dot{\varphi}^{2}}{2(1-{{\cal{T}}}\dot{\varphi}^{2})^{\frac{3}{2}}}\,,
\end{eqnarray}
with the following sound speed
\begin{eqnarray}
    \label{eq6} c_{s}^{2}=1-{\cal{T}}\dot{\varphi}^{2}\,.
\end{eqnarray}
Using the background quantities, we define the slow-roll parameters as
\begin{eqnarray}
    \label{eqn10}
    \epsilon = -\frac{\dot{H}}{H^2}, \qquad \eta = -\frac{1}{H} \frac{\ddot{H}}{\dot{H}}, \qquad s = \frac{1}{H} \frac{\dot{c}_s}{c_s}.\nonumber
\end{eqnarray}

Substituting appropriate expressions, the slow-roll parameter $\epsilon$ becomes
\begin{eqnarray}
    \label{eq7}
    \epsilon=\Bigg[\frac{{\cal{T}}'^{2}V'^{2}}{2\kappa^{2}(1+V{\cal{T}})^{2}}+\frac{{\cal{T}}'^{2}}{2{\cal{T}}^{2}\kappa^{2}(1+V{\cal{T}})^{2}}-
    \frac{{\cal{T}}'V'}{\kappa^{2}(1+V{\cal{T}})^{2}}\Bigg]+\Bigg[\frac{3{\cal{T}}\sum
        (\dot{\zeta}^{j})^{2}}{\kappa^{2}(1+V{\cal{T}})}\Bigg]\equiv
    \epsilon_{DBI}+\epsilon_{Aniso}\,,
\end{eqnarray}
Similarly, $\eta$ is given by

\begin{eqnarray}
    \label{eq8}
    \eta=\Bigg[\frac{{\cal{T}}^{-2}{\cal{T}}''-V''-2{\cal{T}}'^{2}{\cal{T}}^{-3}}{\kappa^{2}({\cal{T}}^{-1}+V)}
    -\frac{\Big(\frac{{\cal{T}}'}{{\cal{T}}^{2}}-V'\Big)^{2}}{2\kappa^{2}({\cal{T}}^{-1}+V)}\Bigg]+\Bigg[\frac{1}{3\kappa^{2}(\frac{{\cal{T}}'}{{\cal{T}}^{2}}-V')}
    \bigg(\sum
    (\ddot{\zeta}^{j})^{2}+\sum\dot{\zeta}^{j}\dddot{\zeta}^{j}\bigg)\Bigg]\nonumber\\
    \equiv\eta_{DBI}+\eta_{Aniso}\,,
\end{eqnarray}
where the parameters with the "DBI" index correspond to the first
brackets and those with the "Aniso" index correspond to the second
brackets in equations (\ref{eq7}) and (\ref{eq8}). We also have
\begin{eqnarray}
    \label{eq9}
    s=\frac{-1}{2[\frac{\kappa^{2}}{3}({\cal{T}}^{-1}+V)+\frac{1}{6}\sum(\dot{\zeta}^{j})^{2}]^{\frac{1}{2}}}\frac{2{\cal{T}}\dot{\varphi}\ddot{\varphi}
        +{\cal{T}}'\dot{\varphi}^{3}}{1-{\cal{T}}\dot{\varphi}^{2}}\,.
\end{eqnarray}

Assuming the special case ${\cal T}^{-1}(\varphi)=V(\varphi)$, we restrict our attention to a specific subclass of DBI models. This condition is adopted as a reconstruction ansatz that reduces the functional freedom associated with the two independent functions ${\cal T}(\varphi)$ and $V(\varphi)$ and allows the background dynamics to be reconstructed in a closed form. Similar reconstruction approaches, in which the DBI warp factor and the scalar potential are related rather than treated as completely independent functions, have been considered previously in the literature~\cite{Ch08}. By this assumption, the slow-roll parameters take the following forms
\begin{equation}
\label{eq7a}
\epsilon=
\left[
\frac{V'^4}{8\kappa^{2}V^{4}}
+
\frac{3V'^2}{8\kappa^{2}V^{2}}
\right]
+
\left[
\frac{3}{2\kappa^{2}V}
\sum_{j=1}^{3}(\dot{\zeta}^{j})^{2}
\right]
\equiv
\epsilon_{\rm DBI}
+
\epsilon_{\rm Aniso}\,,
	\end{equation}	
\begin{equation}
	\label{eq8a}
	\eta=
	\left[
	-\frac{V''+V'^2}{\kappa^{2}V}
	\right]
	-
	\frac{1}{6\kappa^{2}V'}\left[
	\displaystyle\sum_{j=1}^{3}(\ddot{\zeta}^{j})^{2}
		+
		\displaystyle\sum_{j=1}^{3}\dot{\zeta}^{j}\dddot{\zeta}^{j}
	\right]
	\equiv
	\eta_{\rm DBI}
	+
	\eta_{\rm Aniso}\,,
\end{equation}
and
\begin{equation}
	\label{eq9a}
	s=
	-\frac{1}
	{
		2\left[
		\frac{2\kappa^{2}V}{3}
		+
		\frac{1}{6}\sum_{j=1}^{3}(\dot{\zeta}^{j})^{2}
		\right]^{1/2}
	}
	\,
	\frac{
		\frac{2\dot{\varphi}\ddot{\varphi}}{V}
		-
		\frac{V'}{V^{2}}\dot{\varphi}^{3}
	}
	{
		1-\frac{\dot{\varphi}^{2}}{V}
	}\,.
\end{equation}

Also, by assuming ${\cal T}^{-1}(\varphi)=V(\varphi)$, the equations \ref{eq3}-\ref{eq6} yield the following expressions
\begin{equation}
	\label{eq10}
	V=
	\frac{
		\left(
		3H^{2}
		-\frac{1}{2}\sum_{j=1}^{3}(\dot{\zeta}^{j})^{2}
		\right)
		\left(
		3H^{2}
		+2\dot{H}
		+\frac{1}{2}\sum_{j=1}^{3}(\dot{\zeta}^{j})^{2}
		\right)
	}
	{
		2\kappa^{2}\left(3H^{2}+\dot{H}\right)
	}\,
\end{equation}
and
\begin{equation}
	\label{eq11}
	\dot{\varphi}
	=
	\pm
	\frac{1}{\kappa}
	\sqrt{
		\frac{
			\left(
			3H^{2}
			+
			2\dot{H}
			+
			\frac{1}{2}\sum_{j=1}^{3}(\dot{\zeta}^{j})^{2}
			\right)
			\left(
			-2\dot{H}
			-
			\sum_{j=1}^{3}(\dot{\zeta}^{j})^{2}
			\right)
		}
		{
			3H^{2}
			-
			\frac{1}{2}\sum_{j=1}^{3}(\dot{\zeta}^{j})^{2}
		}
	}\,.
\end{equation}
As noted in~\cite{Noj22}, the anisotropy parameters satisfy
\begin{eqnarray}
    \label{eq12}\ddot{\zeta}^{j}+3\,H\,\dot{\zeta}^{j}=0\,.
\end{eqnarray}
whose solution gives
\begin{eqnarray}
    \label{eq13}\dot{\zeta}^{j}=\frac{C^{j}}{a^{3}}\,,
\end{eqnarray}
where $C^j$ are constants constrained by $\sum C^j = 0$ due to the traceless condition $\sum \dot{\zeta}^j = 0$. The nontrivial anisotropic solutions exist whenever at least one of the constants $C^{j}$ is nonvanishing. In this case, the anisotropic sector contributes explicitly to the modified Friedmann equations through the positive-definite term $\sum_{j=1}^{3}(\dot{\zeta}^{j})^{2}$. Hence, anisotropy arises directly from the gravitational dynamics of the anisotropic background geometry and does not require the introduction of an additional vector-field sector or a non-minimal scalar-vector coupling.

With the above background established, we are now ready to examine the perturbative structure of this anisotropic DBI model using the ADM formalism.

\section{\label{s3}Primordial Perturbations}
To analyze scalar perturbations, we employ the ADM formalism in which spacetime is decomposed into a $3+1$ structure. The line element is written as
\begin{eqnarray}
    \label{eq14}ds^2=-N^{2}\,dt^{2}+h_{ij}(dx^{j}+N^{j}\,dt)(dx^{j}+N^{j}\,dt)\,,
\end{eqnarray}
where \( N \) and \( N^i \) are the lapse and shift functions,
respectively. The spatial metric \( h_{ij} \), in the comoving gauge
and within an anisotropic background, is given by
\begin{eqnarray}
    \label{eq15}h_{ij}=a^{2}\,e^{2{\cal{R}}}\,e^{2\zeta_{i}}\,\delta_{ij}\,,
\end{eqnarray}
with \( a \) denoting the scale factor, \( \mathcal{R} \) the
curvature perturbation, and \( \zeta_j \) the anisotropy parameters
(as defined earlier). In the comoving gauge, we set \( \delta\varphi
= 0 \).

Substituting this metric into the action \eqref{eq1}, and setting \( \kappa = 1 \) for simplicity, the action becomes
\begin{eqnarray}
	\label{eq16}
	S
	=
	\int dt\,d^{3}x\,\sqrt{h}
	\bigg[
	\frac{N}{2}R^{(3)}
	+
	\frac{1}{2N}
	\left(E_{ij}E^{ij}-E^{2}\right)
	-
	N{\cal T}^{-1}(\varphi)
	\sqrt{
		1-X\,{\cal T}(\varphi)
	}
	-
	NV(\varphi)
	\bigg] \,
\end{eqnarray}
where \( R^{(3)} \) is the Ricci scalar computed from \( h_{ij} \), \( X = -\partial_\mu \varphi\, \partial^\mu \varphi \) and
\begin{eqnarray}
    \label{eq17}E_{ij}\equiv
    \frac{1}{2}\Big(\dot{h}_{ij}-\nabla_{i}N_{j}-\nabla_{j}N_{i}\Big)\,.
\end{eqnarray}
In the ADM decomposition, the lapse
function $N$ and the shift vector $N_{i}$ enter the action
algebraically, enforcing constraints as Lagrange multipliers rather
than propagating degrees of freedom. Their equations of motion yield
the momentum and Hamiltonian constraints
\begin{eqnarray}
    \label{eq18}\nabla_{j}\left[N^{-1}\Big(E_{i}^{j}-\delta_{i}^{j}E\Big)\right]=0\,,
\end{eqnarray}
\begin{eqnarray}
    \label{eq19}R^{(3)}-\frac{{\cal{T}}^{-1}}{\sqrt{1-{\cal T}X}}+2V-\frac{1}{N^{2}}\bigg(E_{ij}\,E^{ij}-E^{2}\bigg)=0\,.
\end{eqnarray}

Assuming the decomposition \( N = 1 + N_1 \), \( N_i = \partial_i \psi + N_{iT} \) with \( \partial^i N_{iT} = 0 \), we obtain
\begin{eqnarray}
    \label{eq20}E_{ij}=a^{2}e^{2\zeta_{i}}e^{2{\cal{R}}}
    \bigg[\dot{{\cal{R}}}\delta_{ij}+\dot{\zeta}_{i}\delta_{ij}+H\delta_{ij}-\frac{e^{-2\zeta_{i}}e^{-2{\cal{R}}}}{a^{2}}\partial_{i}\partial_{j}\psi\bigg]\,.
\end{eqnarray}
Solving the constraint equations to linear order yields
\begin{eqnarray}
	\label{eq21}
	N_1=\frac{\dot{\cal R}}{H},
	\qquad
	N_{Ti}=0,
	\qquad
	\psi
	=
	-\frac{\cal R}{H}
	+
	\frac{a^2(\epsilon_{\rm DBI}+\epsilon_{\rm Aniso})}{c_s^2}
	\left(
	\sum_{i=1}^{3}e^{-2\zeta_i}\partial_i^2
	\right)^{-1}
	\dot{\cal R}\,.
\end{eqnarray}
Since \( \sum_i \dot{\zeta}_i = 0 \), the lapse function reduces to the familiar isotropic form. Substituting into the action, the second-order action becomes
\begin{eqnarray}
    \label{eq22}S^{(2)}=\int
    dt\,d^{3}x\,a^{3}\,{\cal{G}}\bigg[\dot{{\cal{R}}}^{2}
    -\frac{c_{s}^{2}}{a^{2}}\sum_{i} e^{-2\zeta_{i}}(\partial_{i}{\cal{R}})^{2}\bigg]\,,
\end{eqnarray}
where
\begin{eqnarray}
    \label{eq23}{\cal{G}}=\frac{\epsilon_{DBI}+\epsilon_{Aniso}}{\,c_{s}^{2}}\,.
\end{eqnarray}
The evolution of the curvature perturbation ${\cal{R}}$ is governed by the differential relation derived below
\begin{eqnarray}
    \label{eq24}\ddot{{\cal{R}}}+\frac{\big(a^3{\cal{G}}\big)^{\cdot}}{a^3{\cal{G}}}\dot{{\cal{R}}}
    -\frac{c_{s}^{2}}{a^{2}}\,\sum_{i}e^{-2\zeta_{i}}\,\partial_{i}^{2}{\cal{R}}=0\,.
\end{eqnarray}

Transforming to Fourier space and using \( k^2 = \sum_{i}e^{-2\zeta_i} k_i^{2} \), the equation becomes
\begin{eqnarray}
    \label{eq25}\ddot{{\cal{R}}}+\frac{\big(a^3{\cal{G}}\big)^{\cdot}}{a^3{\cal{G}}}\dot{{\cal{R}}}
    +\frac{c_{s}^{2}k^{2}}{a^{2}}\,{\cal{R}}=0\,.
\end{eqnarray}

Following canonical quantization, the curvature perturbation is promoted to an operator
\begin{eqnarray}
    \label{eq26}{\cal{R}}_{k}=v_{k}(t)\,a_{k}+v^{*}_{k}(t)\,a^{\dag}_{-k}\,,
\end{eqnarray}
where \( v_k \) is the mode function, and operations \( a_k, a_k^\dagger \) denote the operators associated with particle annihilation and creation, respectively.

Instead of assuming a Bunch-Davies vacuum, we consider a more
general non-BD initial condition in the anisotropic background. As
shown in~\cite{Gan11}, the mode function takes the form
\begin{eqnarray}
    \label{eq27}u_{k}(\tau)=\alpha_{k}\,v_{k}(\tau)+\beta_{k}\,v^{*}_{k}(\tau)=\hspace{3cm}\nonumber\\
    \alpha_{k}\,\frac{i He^{-ic_{s}k\tau}}{2k^{\frac{3}{2}}\sqrt{c_{s}(\epsilon_{DBI}+\epsilon_{Aniso})}}(1+ic_{s}k\tau)
    +\beta_{k}\,\frac{i He^{ic_{s}k\tau}}{2k^{\frac{3}{2}}\sqrt{c_{s}(\epsilon_{DBI}+\epsilon_{Aniso})}}(1-ic_{s}k\tau)\,,
\end{eqnarray}
with \( |\alpha_k|^2 - |\beta_k|^2 = 1 \). The operator expression \eqref{eq26} is then rewritten with \( u_k \) replacing \( v_k \).

Using the two-point correlation function
\begin{eqnarray}
    \label{eq28}\langle
    {\cal{R}}_{\textbf{k}}{\cal{R}}_{\textbf{k}'}\rangle=(2\pi)^{3}\delta^{3}(\textbf{k}+\textbf{k}')\frac{2\pi^{2}}{k^{3}}{\cal{P}}_{s}\,,
\end{eqnarray}
the power spectrum becomes
\begin{eqnarray}
    \label{eq29}{\cal{P}}_{s}(k)=\frac{H^{2}}{8\pi^{2}c_{s}(\epsilon_{DBI}+\epsilon_{Aniso})}\Big(1+2|\beta_{k}|^{2}+2|\alpha_{k}|\,|\beta_{k}|\,\cos
    \theta_{k}\Big)\,,
\end{eqnarray}
with \( \theta_k \) denoting the phase between \( \alpha_k \) and \( \beta_k \). Using the occupation number \( N_k = |\beta_k|^2 \), the above can be recast as
\begin{eqnarray}
    \label{eq30}{\cal{P}}_{s}(k)=\frac{H^{2}}{8\pi^{2}c_{s}(\epsilon_{DBI}+\epsilon_{Aniso})}\Big(1+2N_{k}+2\sqrt{N_{k}(1+N_{k})\,}\cos
    \theta_{k}\Big)\,,
\end{eqnarray}
In this setting, the spectral tilt is determined through the following relation
\begin{eqnarray}
    \label{eq31}n_{s}=1-2(\epsilon_{DBI}+\epsilon_{Aniso})-(\eta_{DBI}+\eta_{Aniso})-s+\frac{d\ln(1+2N_{k})}{d\ln
        k}\,.
\end{eqnarray}

In the above, we neglect the oscillatory term \( \cos \theta_k \)
due to its rapid variation in \( k \), which averages out in
observations and introduces no measurable signal in current
datasets.

For tensor perturbations, the second-order action reads
\begin{eqnarray}
    \label{eq32}S^{(2)}_{T}=\int
    dt\,d^{3}x\,\frac{a^{3}}{4}\bigg[\dot{h}_{\lambda}^{2}
    -\frac{1}{a^{2}}\sum_{i}e^{-2\zeta_{i}}(\partial_{i}
    h_{\lambda})^{2}\bigg]\,,
\end{eqnarray}
where \( \lambda = +,\times \) label the polarization modes~\cite{Fel11}. The corresponding tensor power spectrum is
\begin{eqnarray}
    \label{eq33}{\cal{P}}_{T}(k)=\frac{2H^{2}}{\pi^{2}}\Big(1+2N_{k}+2\sqrt{N_{k}(1+N_{k})}\,\cos
        \theta_{k}\Big)\,.
\end{eqnarray}
and the tensor spectral index becomes
\begin{eqnarray}
    \label{eq34}n_{T}=-2(\epsilon_{DBI}+\epsilon_{Aniso})+\frac{d\ln(1+2N_{k})}{d\ln
        k}\,.
\end{eqnarray}
Finally, the tensor-to-scalar ratio is given by
\begin{eqnarray}
    \label{eq35}r=\frac{{\cal{P}}_{T}}{{\cal{P}}_{s}}=16(\epsilon_{DBI}+\epsilon_{Aniso})c_{s}\,,
\end{eqnarray}
in which the anisotropic contribution enters effectively through
$\epsilon_{\mathrm{Aniso}}$. The subsequent discussion is devoted
to exploring the non-Gaussianities within this anisotropic
inflationary model, incorporating non-BD initial states.

\section{\label{s4}Non-Gaussianity}

Non-Gaussianities in the anisotropic DBI setup can be studied by
expanding the action up to third order in the curvature
perturbation.
By expanding we find

\begin{eqnarray}
	\label{eq36}
	S^{(3)}
	=
	\int dt\,d^{3}x\,a^{3}{\cal G}\Bigg[
	-\frac{c_s^2}{a^2}
	{\cal R}
	\sum_i e^{-2\zeta_i}(\partial_i{\cal R})^2
	-\frac{(\Sigma+2\lambda)c_s^2\dot{\cal R}^{\,3}}
	{(\epsilon_{\rm DBI}+\epsilon_{\rm Aniso})H^3}
	+3{\cal R}\dot{\cal R}^{\,2}
	\nonumber\\
		+\frac{c_s^2}{2a^4(\epsilon_{\rm DBI}+\epsilon_{\rm Aniso})}
	\bigg(3{\cal R}-\frac{\dot{\cal R}}{H}\bigg)
	\bigg[
	\sum_{i,j}e^{-2\zeta_i}e^{-2\zeta_j}
	(\partial_i\partial_j\psi)^2
	-
	\bigg(
	\sum_i e^{-2\zeta_i}\partial_i^2\psi
	\bigg)^2
	\bigg]
	\nonumber\\
	-2\frac{c_s^2}{a^4(\epsilon_{\rm DBI}+\epsilon_{\rm Aniso})}
	\bigg(
	\sum_i e^{-2\zeta_i}\partial_i\psi\,\partial_i{\cal R}
	\bigg)
	\bigg(
	\sum_j e^{-2\zeta_j}\partial_j^2\psi
	\bigg)
	\Bigg] \,,
\end{eqnarray}
where the coefficients \( \Sigma \) and \( \lambda \) are defined as
\begin{eqnarray}
    \label{eq37}\Sigma=X\frac{dP}{dX}+2X^{2}\frac{d^{2}P}{dX^{2}}\,,
\end{eqnarray}
\begin{eqnarray}
    \label{eq38}\lambda=X^2\frac{d^{2}P}{dX^{2}}+\frac{2}{3}X^{3}\frac{d^{3}P}{dX^{3}}\,.
\end{eqnarray}

To calculate the bispectrum, we adopt the interaction picture and
identify the third-order Lagrangian as \( \mathcal{L}_3 \), leading
to the interaction Hamiltonian \( \mathcal{H}_{\text{int}} =
-\mathcal{L}_3 \). Employing the in-in formalism, one can evaluate
the three-point function associated with curvature perturbations
through
\begin{eqnarray}
    \label{eq39}\langle \mathcal{R}(\mathbf{k}_1)
    \mathcal{R}(\mathbf{k}_2) \mathcal{R}(\mathbf{k}_3) \rangle = -i
    \int_{-\infty}^{0} d\tau \, a \, \langle [\mathcal{R}(\tau_f,
    \mathbf{k}_1) \mathcal{R}(\tau_f, \mathbf{k}_2) \mathcal{R}(\tau_f,
    \mathbf{k}_3), \mathcal{H}_{\mathrm{int}}(\tau)] \rangle\,,
\end{eqnarray}
where the expectation value is evaluated in a non-vacuum initial
state in the presence of the anisotropic background.

Incorporating the non-Bunch-Davies (non-BD) initial condition via the mode
expansion~\eqref{eq26} and using the modified mode
function~\eqref{eq27}, one obtains the bispectrum as
\begin{eqnarray}
    \label{eq40}\langle \mathcal{R}(\boldsymbol{k}_1)
    \mathcal{R}(\boldsymbol{k}_2) \mathcal{R}(\boldsymbol{k}_3) \rangle
    =(2\pi)^7 \delta^{(3)}(\boldsymbol{k}_1 + \boldsymbol{k}_2 +
    \boldsymbol{k}_3)\frac{\big({\cal{P}}_{s}\big)^{2}}{\prod_{i}k_{i}^{3}}{\cal{F}}\,,
\end{eqnarray}
where the shape function \( \mathcal{F} \) is expressed as
\begin{eqnarray}
    \label{eq41}{\cal{F}}= \Bigg[\bigg( \frac{1}{c_s^2} - 1 -
    \frac{2\lambda}{\Sigma} \bigg) \frac{3k_1^2 k_2^2 k_3^2}{2K^3} +
    \bigg( \frac{1}{c_s^2} - 1 \bigg) \bigg( -\frac{1}{K} \sum_{i>j}
    k_i^2 k_j^2 + \frac{1}{2K^2} \sum_{i\neq j} k_i^2 k_j^3 +
    \frac{1}{8} \sum_{i} k_i^3 \bigg)\Bigg]\,{\cal{M}}\,,
\end{eqnarray}
with \( K = k_1 + k_2 + k_3 \). The modulation factor \( \mathcal{M} \) encodes non-vacuum initial state effects and by using the relations \( N_k = |\beta_k|^2 \) and \( |\alpha_k|^2 - |\beta_k|^2 = 1 \), it is given by
\begin{eqnarray}
    \label{eq43} {\cal{M}} = 1 + \sum_i 2N_{k_i} + \sum_{i} 2N_{k_i}(1
    + N_{k_i}) \cos(2\theta_{k_i}) + 8 \prod_i \sqrt{N_{k_i}(1 +
        N_{k_i})} \cos(\theta_{k_1} + \theta_{k_2} + \theta_{k_3}),
\end{eqnarray}
where \( \theta_{k_i} \) is the phase of \( \beta_{k_i} \). The non-linearity parameter in this setup is defined by
\begin{eqnarray}
    \label{eq44}
    f_{NL}=\frac{10{\cal{F}}}{3\sum_{i=1}^{3}k_{i}^{3}}{\cal{C}}^{-2}\,,
\end{eqnarray}
with
\begin{eqnarray}
    \label{eq45} {\cal{C}}=1+2N_{k}+2\sqrt{N_k(1+N_k)}\cos\theta_k\,.
\end{eqnarray}

In the equilateral limit where \( k_1 = k_2 = k_3 \), the expression for \( f_{\mathrm{NL}} \) becomes
\begin{eqnarray}
    \label{eq46}
    f_{NL}^{equil}=\Bigg[\frac{85}{324}\bigg(1-\frac{1}{c_{s}^{2}}\bigg)-\frac{10}{81}\frac{\lambda}{\Sigma}\Bigg]\frac{{\cal{M}}^{equil}}{({\cal{C}}^{equil})^{2}}\,.
\end{eqnarray}

The equilateral configuration is of particular interest as DBI
inflation predicts its maximal non-Gaussian signal in this limit.
Having studied both linear and nonlinear perturbations, we now
proceed to numerical analysis. By applying specific ansatze for the
scale factor and initial particle occupation number, we explore the
parameter space in light of current observational constraints.

\section{\label{s5}Numerical Analysis and Observational Viability}
In order to carry out the numerical investigation of our model, we adopt an intermediate form for the scale factor expressed as
\begin{eqnarray}
    \label{eq47} a=a_{0}e^{bt^{\beta}}\,,
\end{eqnarray}
where \(a_0\), \(b\), and \(\beta\) are positive constants, with the
parameter \(\beta\) confined to the interval \(0<\beta<1\). This
functional form describes a growth rate between power-law and
exponential expansions. The intermediate expansion scenario is adopted here as a phenomenological background ansatz rather than as a unique prediction of the anisotropic DBI model. This background interpolates between power-law and exponential expansion and provides a convenient framework for investigating the evolution of the anisotropic sector. Since the anisotropy variables satisfy Eq.~(\ref{eq12}), implying $\dot{\zeta}^{i}\propto a^{-3}$, the dilution of anisotropy is directly controlled by the background expansion history. Compared with exponential expansion, the intermediate scenario allows anisotropic contributions to remain dynamically relevant for a longer period while still yielding accelerated expansion. Consequently, it offers a useful setting for exploring potential observational signatures of small anisotropic corrections. Based on this scale factor, the Hubble parameter \(H = \dot{a}/a\) can be written in terms of the number of e-folds \({\cal N} = \int H\, dt\) as
\begin{eqnarray}
    \label{eq48}
    H={\cal{N}}\,\beta\,\left(\frac{{\cal{N}}}{b}\right)^{-\frac{1}{\beta}}\,.
\end{eqnarray}
Furthermore, using the background dynamics, the time evolution of the anisotropic parameter \(\zeta^i\) can be derived as
\begin{eqnarray}
    \label{eq49}
    \dot{\zeta}^{i}=\frac{C^{i}}{a_{0}^{3}e^{3{\cal{N}}}}\,,
\end{eqnarray}
with \(C^i\) being integration constants associated with each spatial direction. From now on, we have $c^{2}=\sum_{i=1}^{3}(C^{i})^{2}$.

These expressions allow us to obtain the slow-roll parameters in
this background as functions of the model parameters. However, due
to the complexity of the resulting expressions, their explicit forms
are deferred to the Appendix. In addition, we adopt a
phenomenological ansatz for the number density of excited modes
following~\cite{Gan11}
\begin{eqnarray}
    \label{eq50} N_{k}=N_{k,0}\,e^{-\frac{k^{2}}{k_{cut}^{2}}}\,,
\end{eqnarray}
where \(k_{\text{cut}}\) denotes a cutoff scale beyond which the
excitation becomes negligible. To ensure consistency with the
inflationary background, the backreaction of excited modes must be
sufficiently suppressed. This leads to the approximate bound
\[
N_{k,0} \lesssim \frac{M_{\text{pl}}^2 H^2}{k_{\text{cut}}^4}\,,
\]
as discussed in~\cite{Gan11}. With the above form for \(N_k\), the scalar spectral index \(n_s\) acquires an additional scale-dependent correction and is given by
\begin{eqnarray}
    \label{eq51}
    n_{s}-1=-2(\epsilon_{DBI}+\epsilon_{Aniso})-(\eta_{DBI}+\eta_{Aniso})-s
    -\frac{4N_{k,0}\,k^{2}}{k_{cut}^{2}\Big(1+2N_{k,0}e^{-\frac{k^{2}}{k_{cut}^{2}}}\Big)}e^{-\frac{k^{2}}{k_{cut}^{2}}}\,.
\end{eqnarray}

Equipped with the necessary dynamical and perturbative equations, we
proceed to investigate the phenomenological viability of our setup
through numerical analysis. Among various inflationary signatures,
particular attention is paid to the scalar spectral index $n_{s}$
and the tensor-to-scalar ratio $r$, both of which are tightly
constrained by cosmological observations.

According to the Planck 2018 data release, which includes
temperature (TT), E-mode polarization (TE and EE), low-multipole
polarization (low-\(\ell\)), lensing, baryon acoustic oscillations
(BAO), and BICEP2/Keck Array 2018 (BK18) datasets, the upper bound
on the tensor-to-scalar ratio is
\[
r < 0.036 \quad \text{(95\% CL, within the } \Lambda\text{CDM}+r+\frac{dn_s}{d\ln k} \text{ model)}~\cite{Pa22}\,.
\]
In the same analysis, the scalar spectral index is measured to be
\(n_s = 0.9658 \pm 0.0038\). In addition, the Dark Energy
Spectroscopic Instrument (DESI) collaboration has recently released
new data~\cite{Ad25,Ad24}, which further sharpen these constraints.
For instance, as reported in~\cite{Wa24}, the joint analysis of
BK18, CMB, and DESI data yields
\[
n_s = 0.9700 \pm 0.0036\,, \quad r = 0.0176^{+0.0070}_{-0.0130}\,.
\]
A more comprehensive survey of inflationary constraints
incorporating various datasets has also been conducted
in~\cite{Co24}. Within the standard \(\Lambda\)CDM framework,
several combinations of data lead to the following sample bounds
\begin{itemize}
    \item SDSS + CMB + Union3: \(n_s = 0.9652 \pm 0.0037\), \(r = 0.0172^{+0.0071}_{-0.013}\)
    \item DESI + CMB + Union3: \(n_s = 0.9673 \pm 0.0036\), \(r = 0.0178^{+0.0077}_{-0.013}\)
    \item SDSS + CMB + DESY5: \(n_s = 0.9644 \pm 0.0036\), \(r = 0.0169^{+0.0069}_{-0.013}\)
    \item DESI + CMB + DESY5: \(n_s = 0.9664 \pm 0.0035\), \(r = 0.0176^{+0.0073}_{-0.013}\)
\end{itemize}
These observational constraints offer a benchmark for testing the
validity of any inflationary scenario, including our anisotropic DBI
model with non-standard initial conditions.

\begin{figure}[htbp]
\centering
\includegraphics[width=0.45\textwidth]{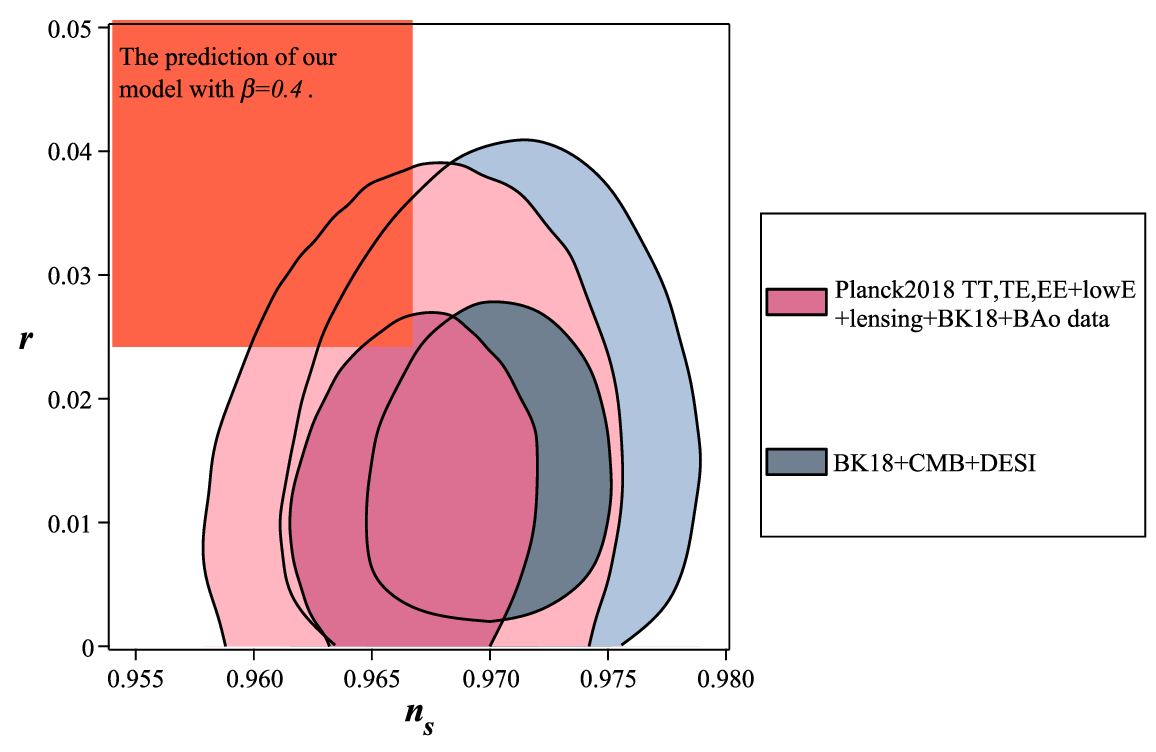}
\includegraphics[width=0.45\textwidth]{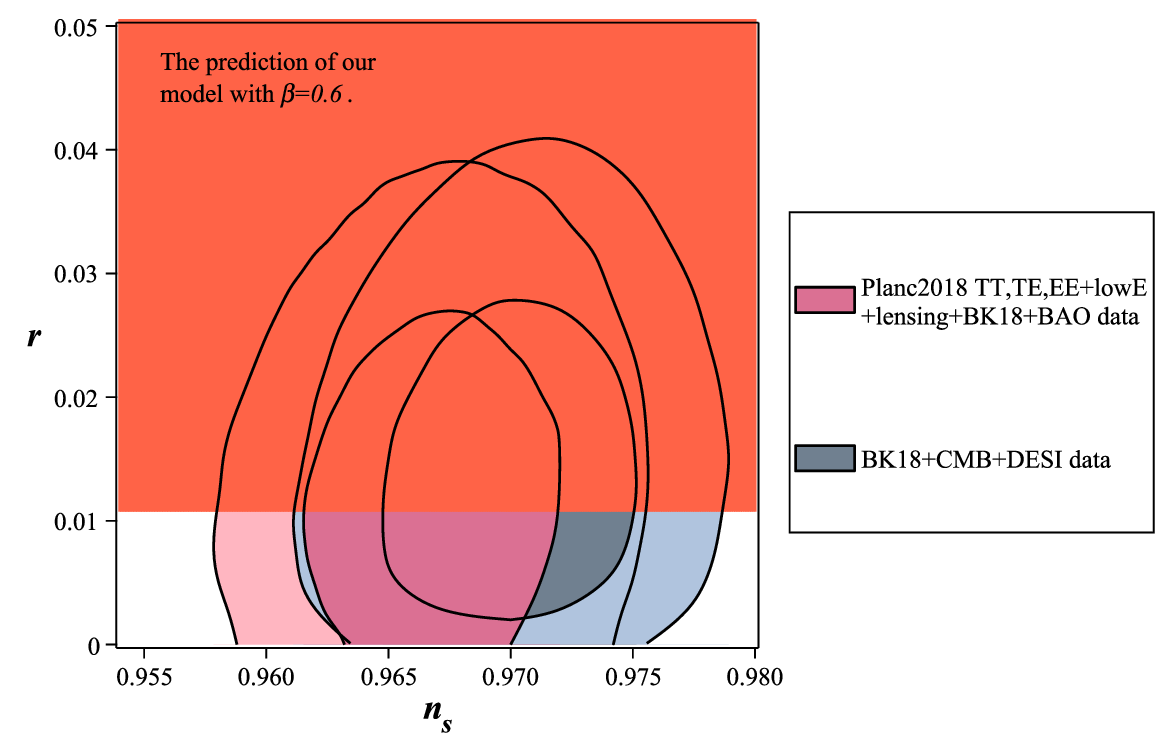}
\includegraphics[width=0.45\textwidth]{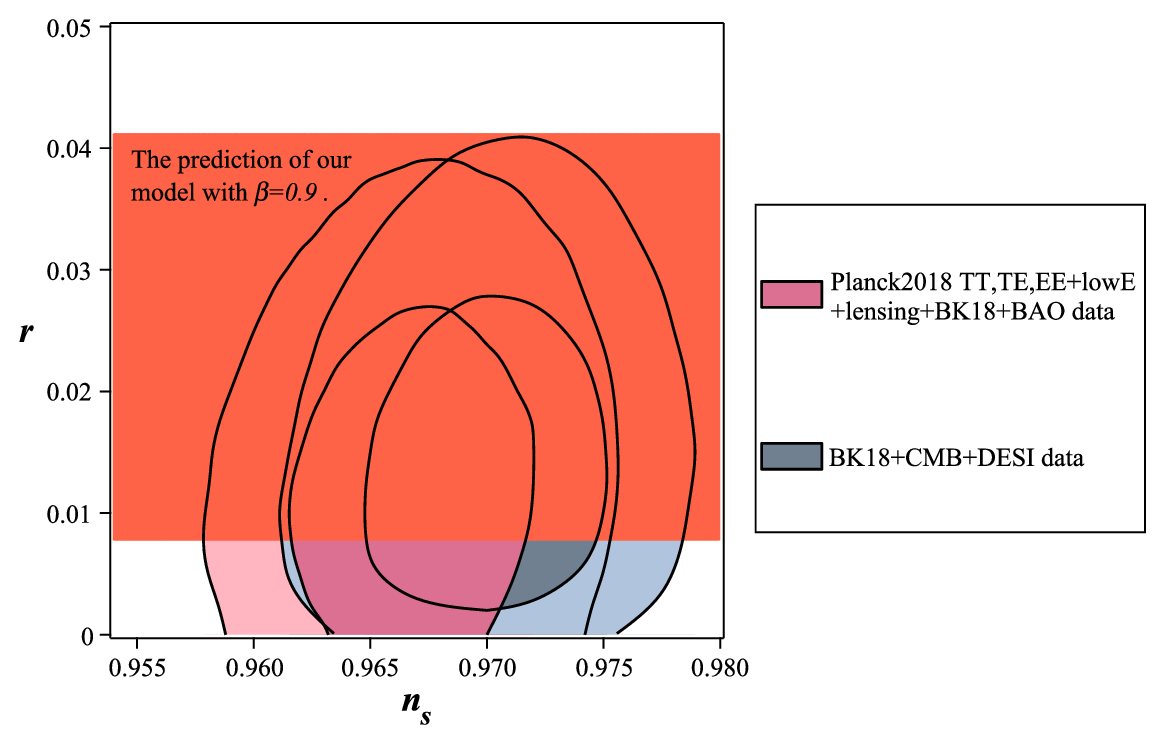} 
\caption{\small
The tensor-to-scalar ratio plotted against the scalar spectral index for the intermediate DBI inflationary scenario with anisotropic geometry, in the
background of Planck2018 TT, TE, EE +lowE+lensing+BK18+BAO data and
DESI+CMB+DESY5 data. The upper left panel is for
$\beta=0.4$, the upper right panel is for $\beta=0.6$ and the lower one is for $\beta=0.9$.} \label{fig1}
\end{figure}

\begin{figure}[htbp]
\includegraphics[width=0.35\textwidth]{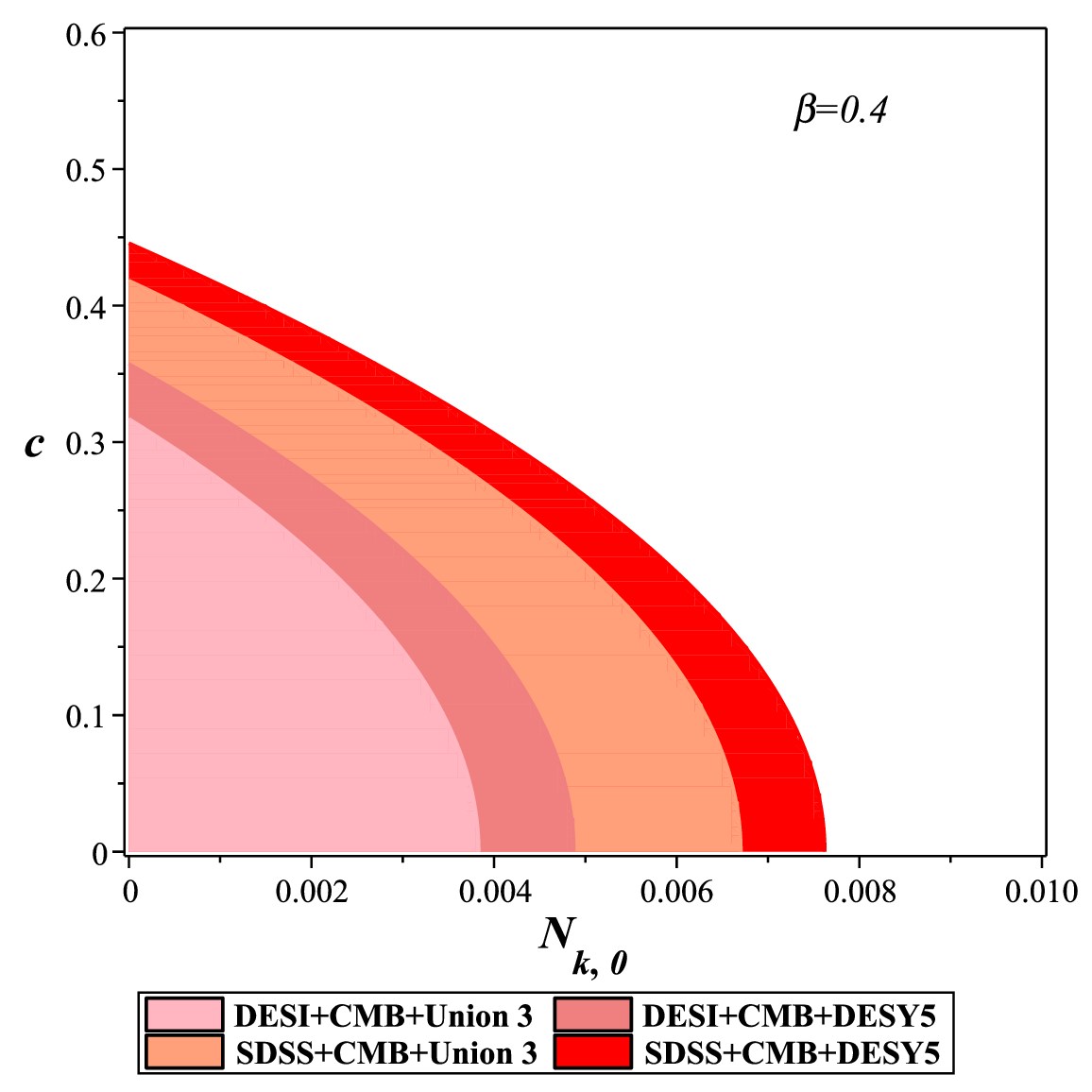}
\includegraphics[width=0.32\textwidth]{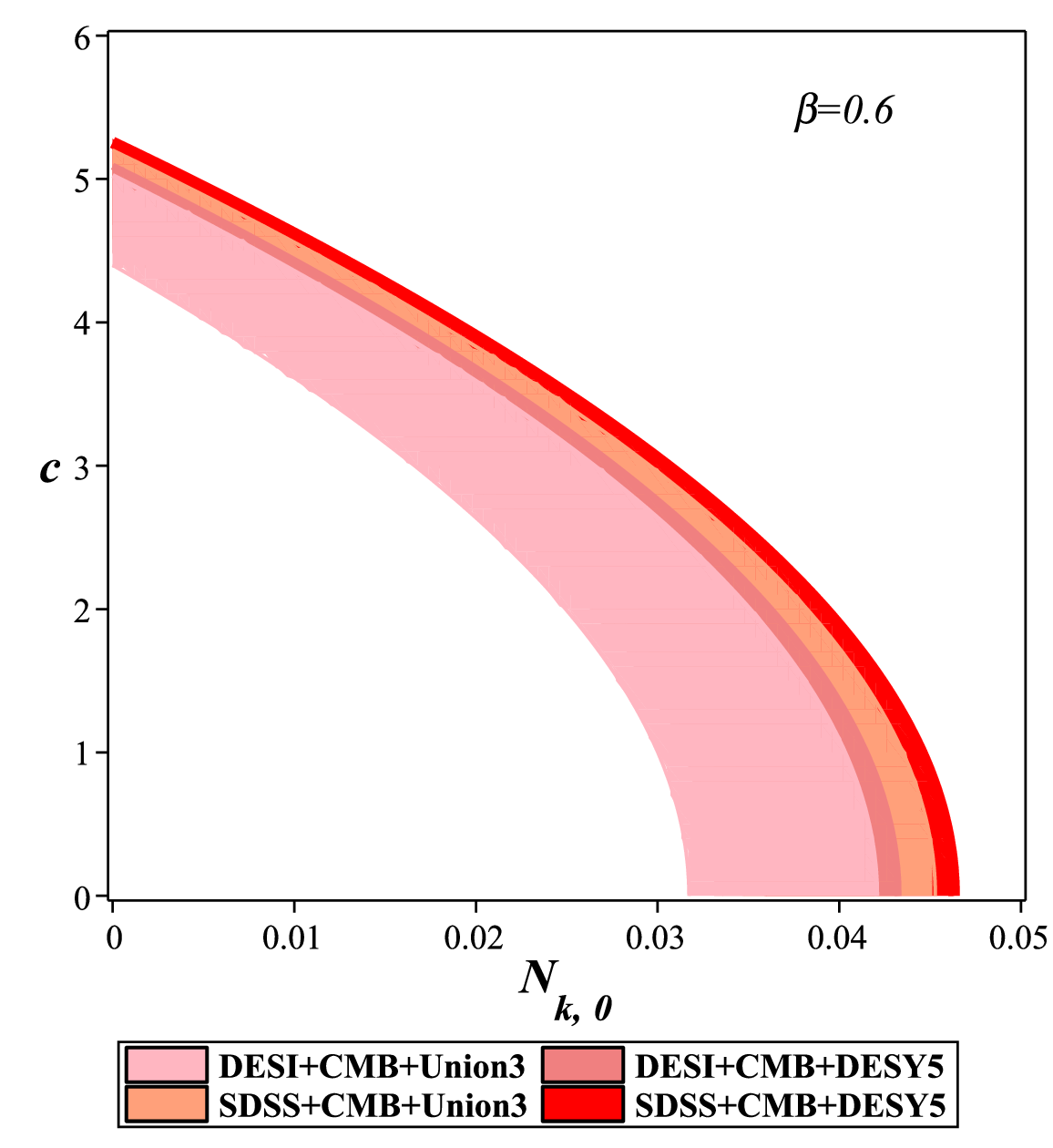} 
\includegraphics[width=0.34\textwidth]{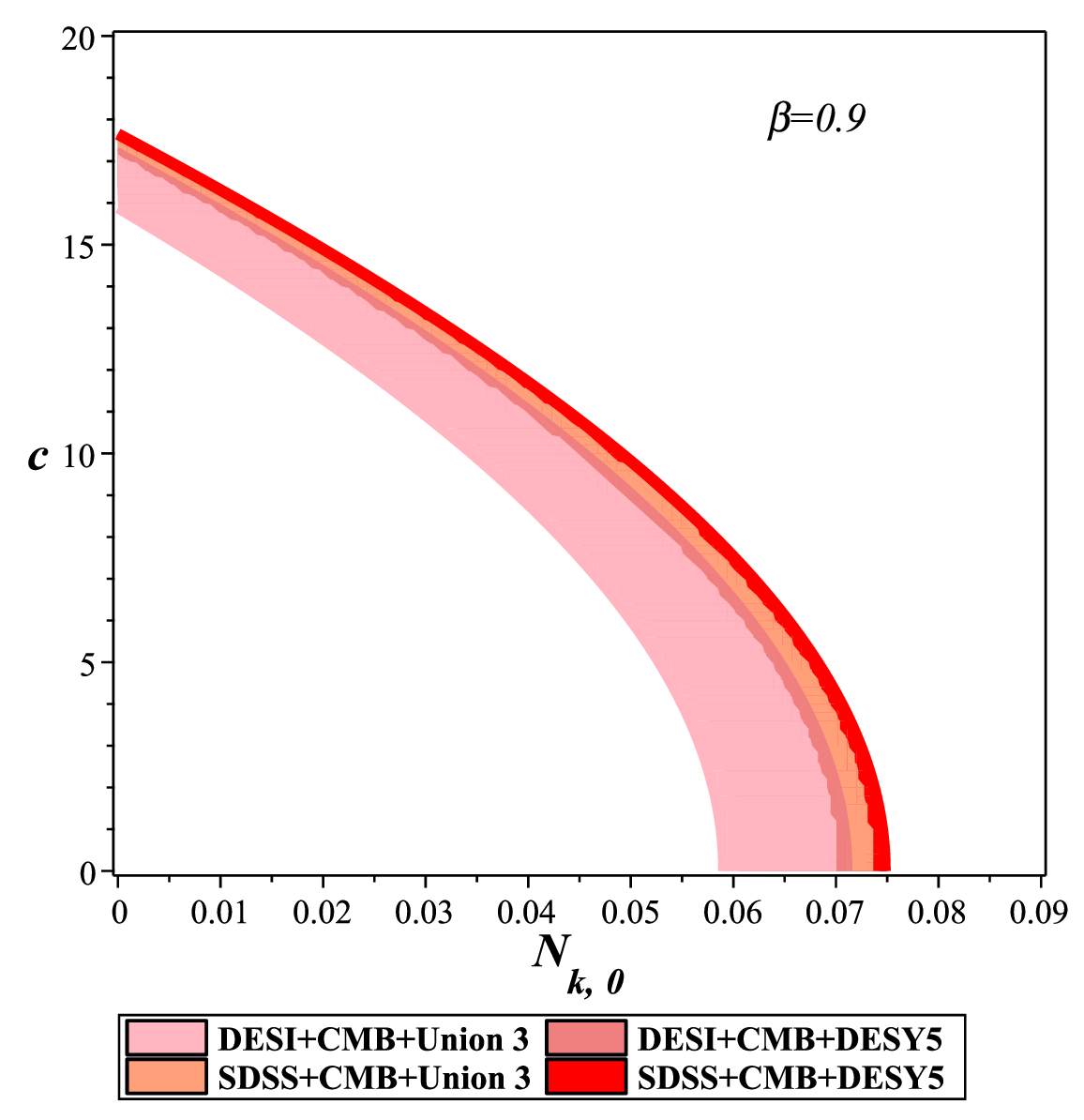} 
\caption{\small Ranges of the model's parameters $c$ and $N_{k,0}$
in which the intermediate DBI model with anisotropic geometry is
consistent with SDSS+CMB+Union3 data, DESI+CMB+Union3 data,
SDSS+CMB+DESY5 data and DESI+CMB+DESY5 data. The upper left panel is for
$\beta=0.4$, the upper right panel is for $\beta=0.6$ and the lower one is for $\beta=0.9$. } \label{fig2}
\end{figure}

To explore the compatibility of our model with current data, we
numerically evaluate the evolution of \(n_s\) and \(r\) for
different values of the parameter \(\beta\), corresponding to three
representative cases: \(\beta = 0.4\), \(\beta = 0.6\), and \(\beta
= 0.9\). We also choose $a_{0}=1$. The resulting trajectories in the
\(r\)--\(n_s\) plane are illustrated in
Figure~\ref{fig1}. In this analysis, the ranges \(0 <
N_{k,0} < 0.1\) and \(0 < c < 30\) have been considered. Our simulations
indicate that certain regions in this parameter space yield
predictions that fall within the allowed observational bounds.
Additionally, to provide a more complete picture, we investigate the
model's behavior against various observational priors by plotting
the \(c\)--\(N_{k,0}\) trajectories under alternative data backgrounds
in Figure~\ref{fig2}. The parameter ranges used for
these plots are consistent with those employed in
Figure~\ref{fig1}.

\begin{figure}[htbp]
\includegraphics[width=0.32\textwidth]{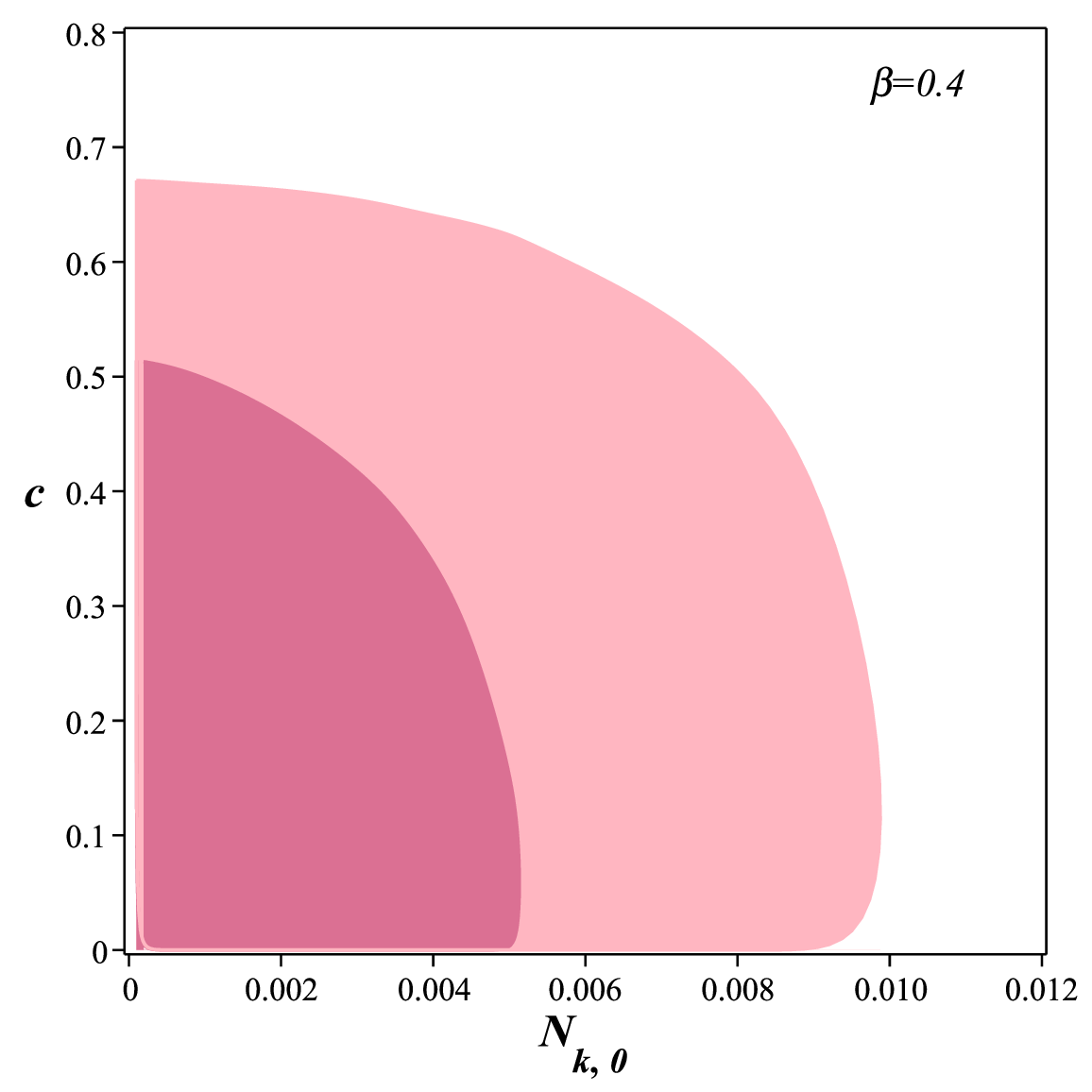} 
\includegraphics[width=0.32\textwidth]{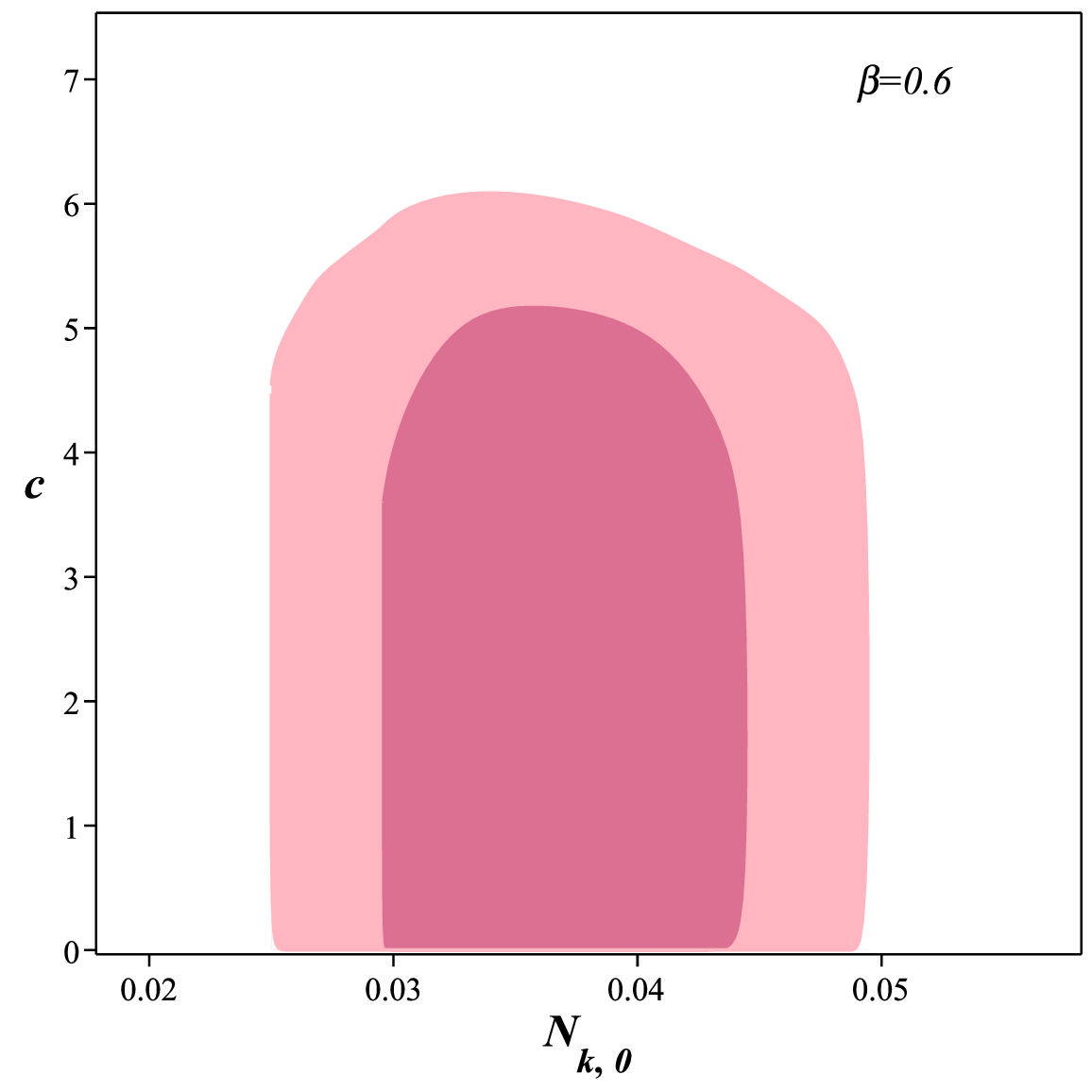} 
\includegraphics[width=0.32\textwidth]{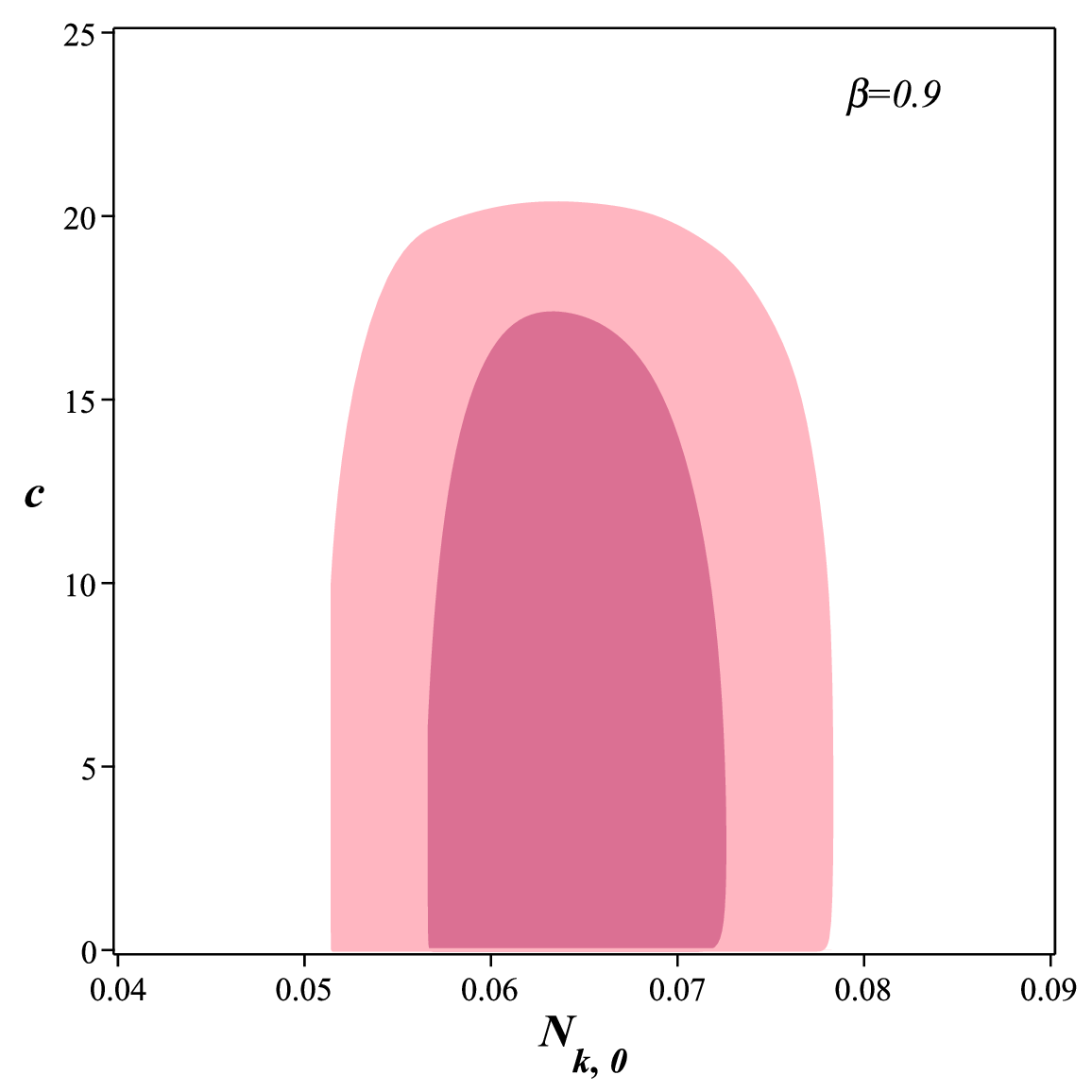}
\caption{\small Domains of the parameters $c$ and $N_{k,0}$ that
make the scalar spectral index and tensor-to-scalar ratio in the
model observationally viable. This figure is based on Planck2018 TT,
TE, EE +lowE+lensing+BK18+BAO data at both $68\%$ and $95\%$ CL.}
\label{fig7}
\end{figure}

\begin{figure}[htbp]
\includegraphics[width=0.32\textwidth]{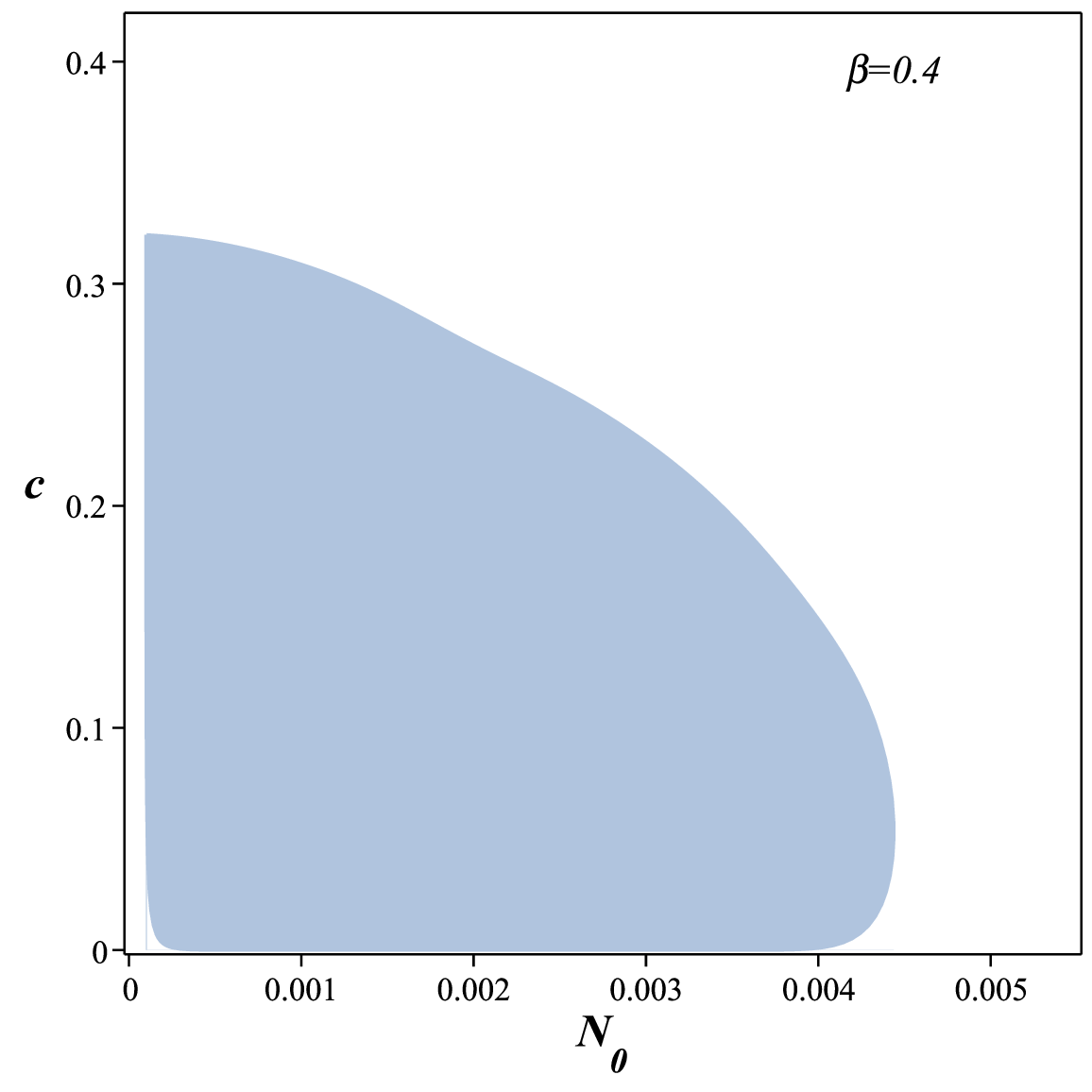} 
\includegraphics[width=0.32\textwidth]{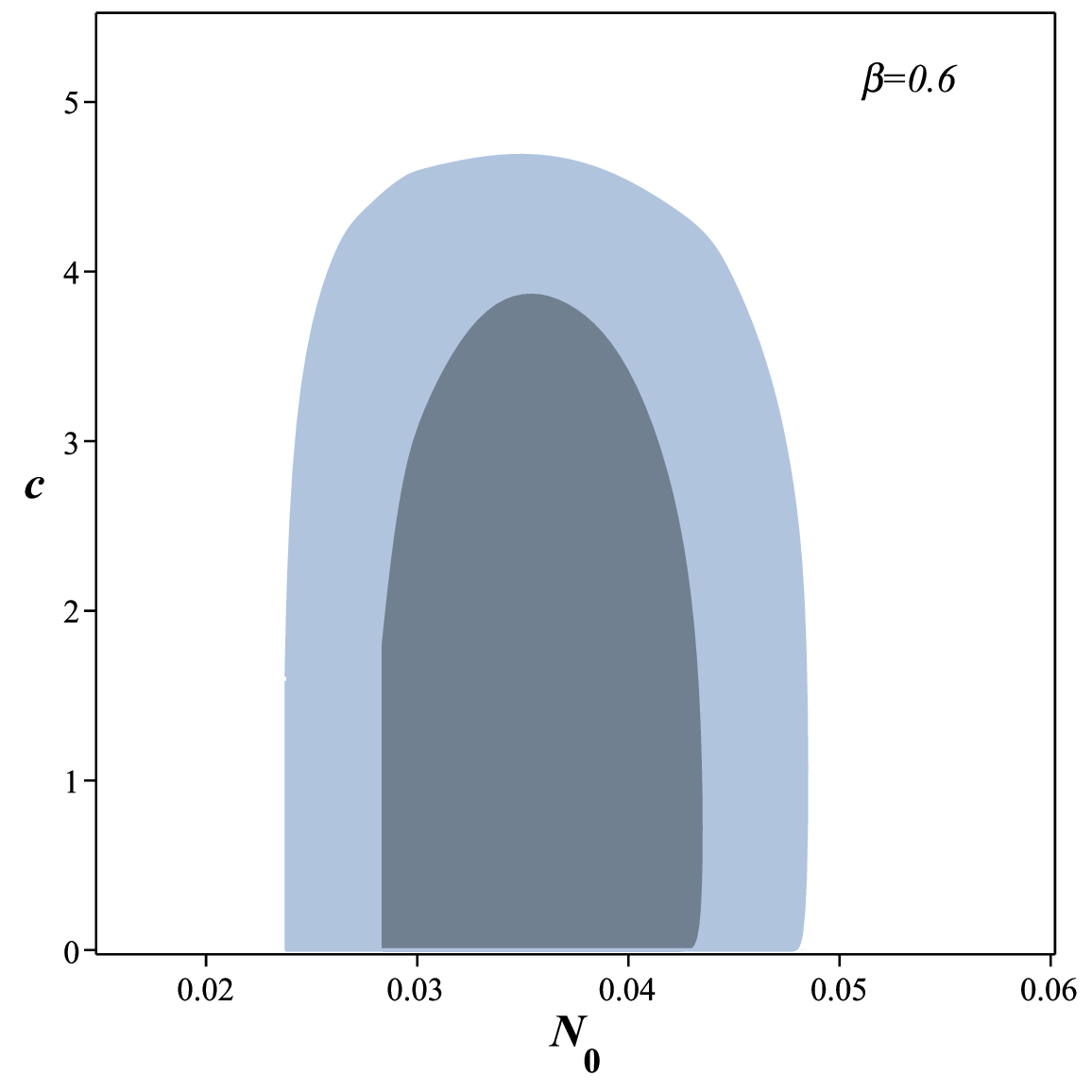} 
\includegraphics[width=0.32\textwidth]{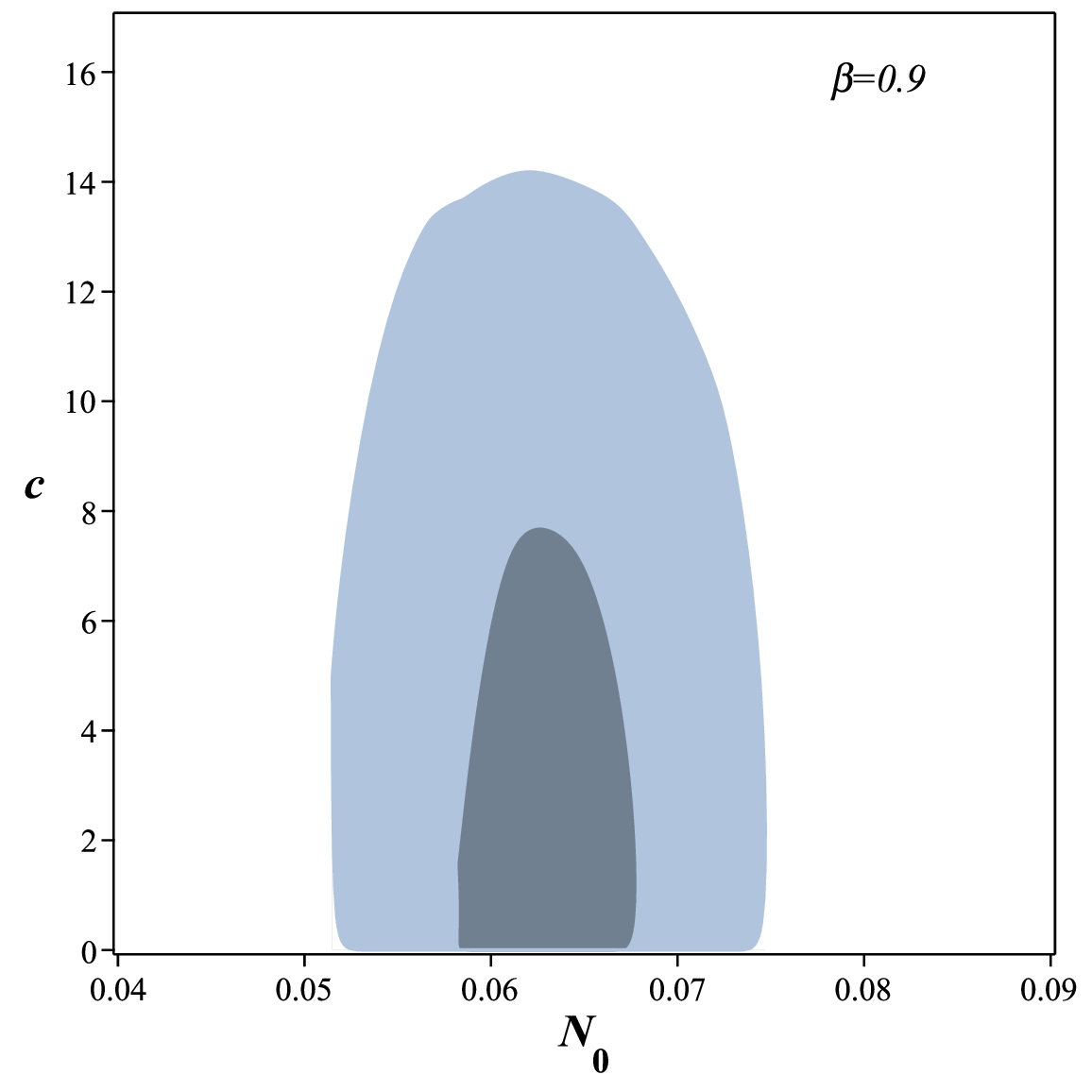}
\caption{\small Domains of the parameters $c$ and $N_{k,0}$ that
make the scalar spectral index and tensor-to-scalar ratio in the
model observationally viable. This figure is based on DESI+CMB+DESY5
data. } \label{fig8}
\end{figure}

To systematically explore the observational viability of our model,
we have carried out numerical investigations based on two
complementary data sets: Planck2018 TT, TE, EE + low-$\ell$ +
lensing + BK18 + BAO, and the DESI+CMB+DESY5 dataset. Our analysis
includes confidence levels at both 68\% and 95\%, and we have
examined three benchmark values for the intermediate expansion
parameter: \(\beta = 0.4\), \(0.6\), and \(0.9\). For the Planck2018
dataset at 95\% CL, our results indicate that the parameter space is
consistent with observations within the ranges:
\[
0 < c < 0.671 \quad \text{and} \quad 1.00 \times 10^{-3} < N_{k,0} < 1.00 \times 10^{-2} \quad \text{for } \beta = 0.4\,,
\]
\[
0 < c < 6.38 \quad \text{and} \quad 2.50 \times 10^{-2} < N_{k,0} < 4.95 \times 10^{-2} \quad \text{for } \beta = 0.6\,,
\]
\[
0 < c < 20.4 \quad \text{and} \quad 5.15 \times 10^{-2} < N_{k,0} < 7.83 \times 10^{-2} \quad \text{for } \beta = 0.9\,.
\]
At 68\% CL, the compatible intervals become slightly narrower:
\[
0 < c < 0.514 \quad \text{and} \quad 1.00 \times 10^{-3} < N_{k,0} < 5.15 \times 10^{-3} \quad \text{for } \beta = 0.4\,,
\]
\[
0 < c < 5.18 \quad \text{and} \quad 3.60 \times 10^{-2} < N_{k,0} < 4.45 \times 10^{-2} \quad \text{for } \beta = 0.6\,,
\]
\[
0 < c < 17.4 \quad \text{and} \quad 5.67 \times 10^{-2} < N_{k,0} < 7.26 \times 10^{-2} \quad \text{for } \beta = 0.9\,.
\]
In a similar fashion, we have repeated the numerical analysis using
the DESI+CMB+DESY5 dataset. For \(\beta = 0.4\), the model is only
compatible with the observational data at 95\% CL, with the allowed
range:
\[
0 < c < 3.22 \quad \text{and} \quad 1.00 \times 10^{-3} < N_{k,0} < 4.50 \times 10^{-3}\,.
\]
For \(\beta = 0.6\), the acceptable regions are
\[
\begin{aligned}
    &0 < c < 4.69 \quad \text{for} \quad 2.38 \times 10^{-2} < N_{k,0} < 4.85 \times 10^{-2} \quad \text{(95\% CL)}\,,\\
    &0 < c < 3.87 \quad \text{for} \quad 2.84 \times 10^{-2} < N_{k,0} < 4.35 \times 10^{-2} \quad \text{(68\% CL)}\,.
\end{aligned}
\]
Lastly, for \(\beta = 0.9\), the viable range at 95\% CL is
\[
0 < c < 14.2 \quad \text{and} \quad 5.15 \times 10^{-2} < N_{k,0} < 7.48 \times 10^{-2}\,.
\]
To visualize these viable regions, we present the allowed domains in
the \(c\)--\(N_{k,0}\) parameter space in Figures~\ref{fig7}
and~\ref{fig8}, based on Planck2018 and DESI datasets, respectively.
Additionally, a detailed breakdown of the allowed parameter values
for selected samples of \(N_{k,0}\) is provided in
Tables~\ref{tab1}--\ref{tab6}.

\begin{table*}[htbp]
\tiny\tiny\caption{\small{\label{tab1} Consistency regions for the
parameter $c$, where the anisotropic DBI model with a non-standard
initial state predicts scalar and tensor observables in agreement
with various data combinations. This table is for $\beta=0.4$}}
\begin{center}
\tabcolsep=0.05cm\begin{tabular}{|c|c|c|c|c|c|} \hline \hline &&&&\\
& Planck2018 TT,TE,EE+lowE & Planck2018 TT,TE,EE+lowE &
DESI+CMB+DESY5 & DESI+CMB+DESY5
\\
& +lensing+BK18+BAO & +lensing+BK18+BAO& &
\\
\hline &&&&\\$N_{k,0}$& $68\%$ CL & $95\%$ CL &$68\%$ CL & $95\%$ CL
\\
\hline\hline &&&&\\  $1.00\times10^{-3}$ & $0<c<0.514$ & $0<c<0.671
$ & --- & $0<c<0.322$\\&&&& \\
\hline &&&&\\$4.00\times10^{-3}$& $0.00<c<0.340$ & $0.00<c<0.640$
&---& $0.00<c<0.150$
\\ &&&& \\ \hline &&&&\\
$7.00\times10^{-3}$& --- & $0.00<c<0.560
$ & --- & --- \\ &&&& \\
\hline &&&&\\
$1.00\times10^{-2}$& --- & $0.00<c<0.210$ & --- &
---
\\ &&&&\\
\hline
\end{tabular}
\end{center}
\end{table*}

\begin{table*}[htbp]
\tiny\tiny\caption{\small{\label{tab2} Consistency regions for the
parameter $c$, where the anisotropic DBI model with a non-standard
initial state predicts scalar and tensor observables in agreement
with various data combinations. This table is for $\beta=0.6$}}
\begin{center}
\tabcolsep=0.05cm\begin{tabular}{|c|c|c|c|c|c|} \hline \hline &&&&\\
& Planck2018 TT,TE,EE+lowE & Planck2018 TT,TE,EE+lowE &
DESI+CMB+DESY5  & DESI+CMB+DESY5
\\
& +lensing+BK18+BAO & +lensing+BK18+BAO& &
\\
\hline &&&&\\$N_{k,0}$& $68\%$ CL & $95\%$ CL &$68\%$ CL & $95\%$ CL
\\
\hline\hline &&&&\\  $3.00\times10^{-2}$ & $0<c<4.21$ & $0<c<5.53
$ & $0<c<3.10$ & $0<c<4.59$\\&&&& \\
\hline &&&&\\$3.50\times10^{-2}$& $0.00<c<5.18$ & $0.00<c<6.09$ &
$0.00<c<3.88$ & $0.00<c<4.69$
\\ &&&& \\ \hline &&&&\\
$4.00\times10^{-2}$& $0.00<c<5.01$ & $0.00<c<5.71 $ & $0.00<c<3.49 $
& $0.00<c<4.54$ \\ &&&& \\
\hline &&&&\\
$4.50\times10^{-2}$& $0.00<c<3.35 $ & $0.00<c<5.38$ & --- &
$0.00<c<4.07 $
\\ &&&&\\
\hline
\end{tabular}
\end{center}
\end{table*}

\begin{table*}[htbp]
\tiny\tiny\caption{\small{\label{tab3} Consistency regions for the
parameter $c$, where the anisotropic DBI model with a non-standard
initial state predicts scalar and tensor observables in agreement
with various data combinations. This table is for $\beta=0.9$}}
\begin{center}
\tabcolsep=0.05cm\begin{tabular}{|c|c|c|c|c|c|} \hline \hline &&&&\\
& Planck2018 TT,TE,EE+lowE & Planck2018 TT,TE,EE+lowE &
DESI+CMB+DESY5 & DESI+CMB+DESY5
\\
& +lensing+BK18+BAO & +lensing+BK18+BAO& &
\\
\hline &&&&\\$N_{k,0}$& $68\%$ CL & $95\%$ CL &$68\%$ CL & $95\%$ CL
\\
\hline\hline &&&&\\  $5.50\times10^{-2}$ & --- & $0<c<19.5
$ & --- & $0<c<12.1$\\&&&& \\
\hline &&&&\\$6.00\times10^{-2}$& $0.00<c<16.7$ & $0.00<c<20.2$ &
$0.00<c<6.00$ & $0.00<c<14.0$
\\ &&&& \\ \hline &&&&\\
$6.50\times10^{-2}$& $0.00<c<17.2$ & $0.00<c<20.3 $ & $0.00<c<7.04 $
& $0.00<c<13.9$ \\ &&&& \\
\hline &&&&\\
$7.00\times10^{-2}$& $0.00<c<14.5 $ & $0.00<c<19.7$ & --- &
$0.00<c<12.0 $
\\ &&&&\\
\hline
\end{tabular}
\end{center}
\end{table*}

\begin{table*}[htbp]
\tiny\tiny\caption{\small{\label{tab4} Consistency regions for the
parameter $c$ where the anisotropic DBI model with a non-standard
initial state predicts scalar and tensor observables in agreement
with various data combinations. This table is for $\beta=0.4$}}
\begin{center}
\tabcolsep=0.05cm\begin{tabular}{|c|c|c|c|c|c|} \hline \hline &&&&\\
& SDSS+CMB+Union3 & DESI+CMB+Union3 & SDSS+CMB+DESY5 &
DESI+CMB+DESY5
\\
\hline &&&&\\$N_{k,0}$& $68\%$ CL & $68\%$ CL &$68\%$ CL & $68\%$ CL
\\
\hline\hline &&&&\\  $1.00\times10^{-3}$ & $0<c<0.383$ & $0<c<0.271
$ & $0<c<0.413$ & $0<c<0.315$\\&&&& \\
\hline &&&&\\$4.00\times10^{-3}$& $0.00<c<0.26$ & --- &
$0.00<c<0.304$ & $0.00<c<0.148$
\\ &&&& \\ \hline &&&&\\
$7.00\times10^{-3}$& --- & --- & $0.00<c<0.122$ & --- \\ &&&& \\
\hline &&&&\\
$1.00\times10^{-2}$& --- & --- & --- &
---
\\ &&&&\\
\hline
\end{tabular}
\end{center}
\end{table*}

\begin{table*}[htbp]
\tiny\tiny\caption{\small{\label{tab5} Consistency regions for the
parameter $c$, where the anisotropic DBI model with a non-standard
initial state predicts scalar and tensor observables in agreement
with various data combinations. This table is for $\beta=0.6$}}
\begin{center}
\tabcolsep=0.05cm\begin{tabular}{|c|c|c|c|c|c|} \hline \hline &&&&\\
& SDSS+CMB+Union3 & DESI+CMB+Union3 & SDSS+CMB+DESY5 &
DESI+CMB+DESY5
\\
\hline &&&&\\$N_{k,0}$& $68\%$ CL & $68\%$ CL &$68\%$ CL & $68\%$ CL
\\
\hline\hline &&&&\\  $3.00\times10^{-2}$ & $1.604<c<2.934$ &
$0.981<c<2.604$ & $1.826<c<3.026$ & $1.302<c<2.741$\\&&&& \\
\hline &&&&\\$3.50\times10^{-2}$& $0.00<c<2.399$ & $0.00<c<1.977$ &
$0.613<c<2.505$ & $0.00<c<2.145$
\\ &&&& \\ \hline &&&&\\
$4.00\times10^{-2}$& $0.00<c<1.696$ & $0.00<c<1.039$ & $0.00<c<1.813$ & $0.00<c<1.327$ \\ &&&& \\
\hline &&&&\\
$4.50\times10^{-2}$&$0.00<c<0.044$ & $0.00<c<0.672$ & $0.00<c<0.624$
&
---
\\ &&&&\\
\hline
\end{tabular}
\end{center}
\end{table*}

\begin{table*}[htbp]
\tiny\tiny\caption{\small{\label{tab6} Consistency regions for the
parameter $c$, where the anisotropic DBI model with a non-standard
initial state predicts scalar and tensor observables in agreement
with various data combinations. This table is for $\beta=0.9$}}
\begin{center}
\tabcolsep=0.05cm\begin{tabular}{|c|c|c|c|c|c|} \hline \hline &&&&\\
& SDSS+CMB+Union3 & DESI+CMB+Union3 & SDSS+CMB+DESY5 &
DESI+CMB+DESY5
\\
\hline &&&&\\$N_{k,0}$& $68\%$ CL & $68\%$ CL &$68\%$ CL & $68\%$ CL
\\
\hline\hline &&&&\\  $5.50\times10^{-2}$ & $5.097<c<8.402$ &
$3.888<c<7.609$ & $5.632<c<8.685$ & $4.572<c<7.852$\\&&&& \\
\hline &&&&\\$6.00\times10^{-2}$& $2.712<c<7.160$ & $0.00<c<6.146$ &
$3.389<c<7.465$ & $1.047<c<6.534$
\\ &&&& \\ \hline &&&&\\
$6.50\times10^{-2}$& $0.00<c<5.658$ & $0.00<c<4.308$ & $0.00<c<6.010$ & $0.00<c<4.883$ \\ &&&& \\
\hline &&&&\\
$7.00\times10^{-2}$&$0.00<c<3.616$ & $0.00<c<0.798$ & $0.00<c<4.101$
& $0.00<c<2.180$
\\ &&&&\\
\hline
\end{tabular}
\end{center}
\end{table*}

\begin{figure}[htbp]
\includegraphics[width=0.32\textwidth]{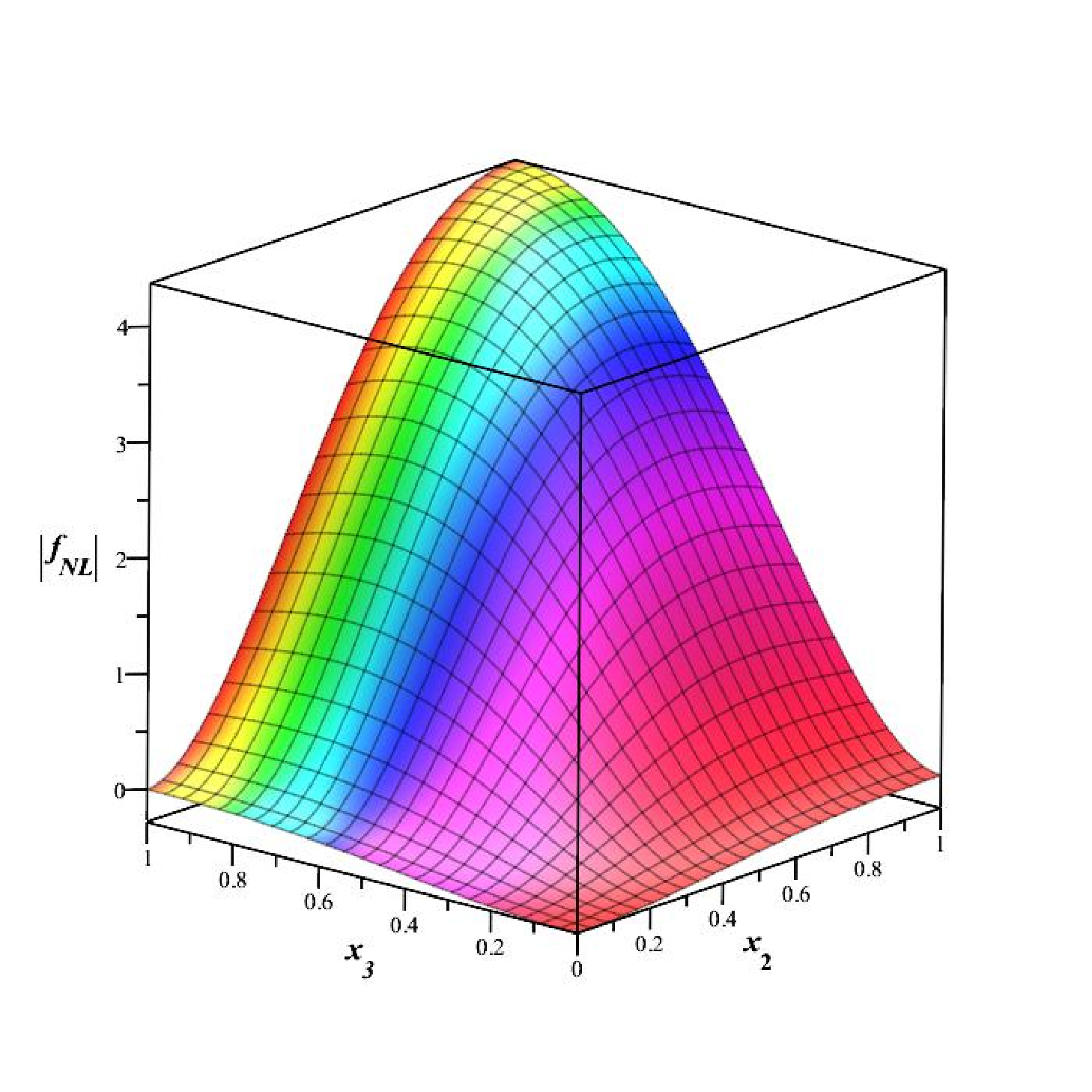} 
\includegraphics[width=0.32\textwidth]{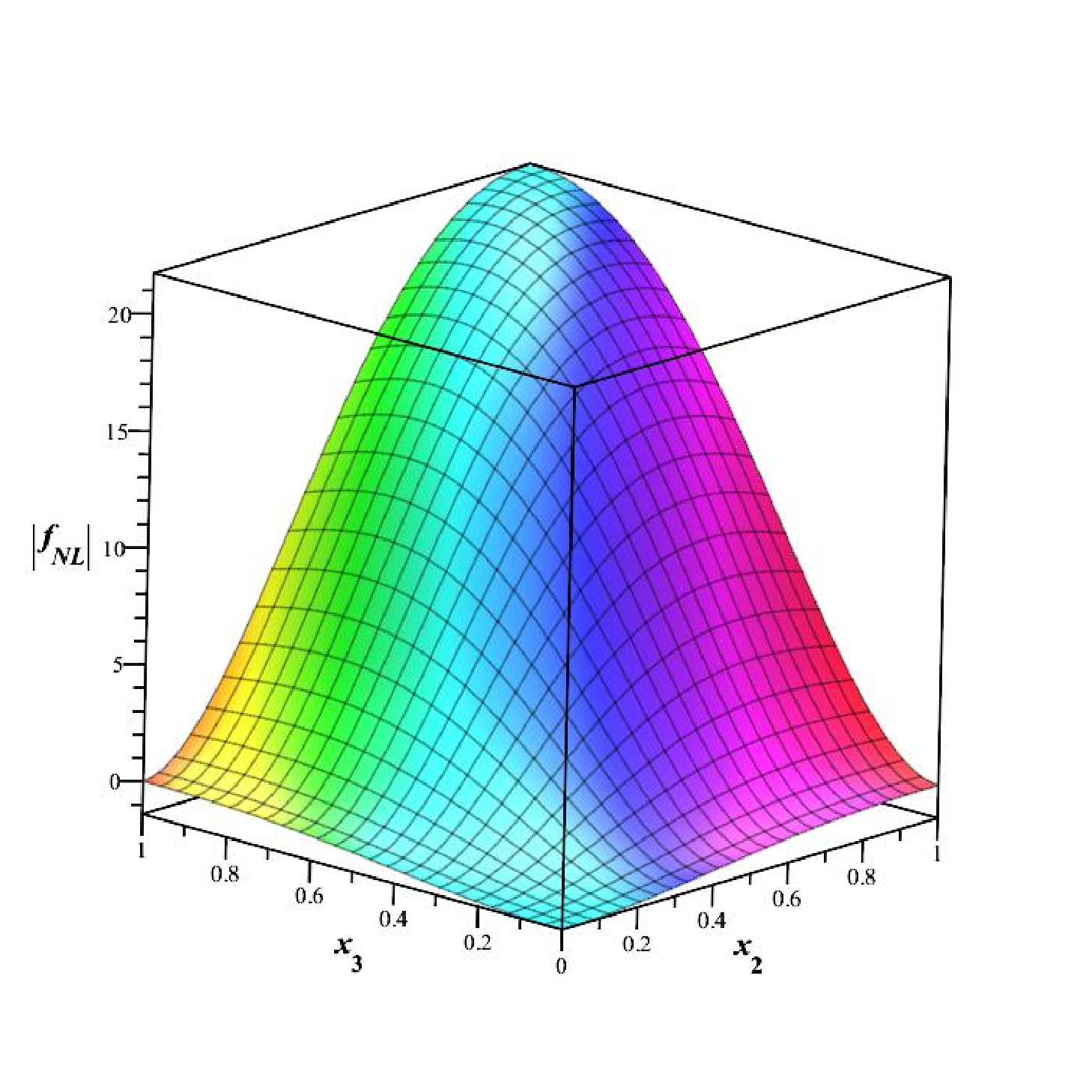} 
\includegraphics[width=0.32\textwidth]{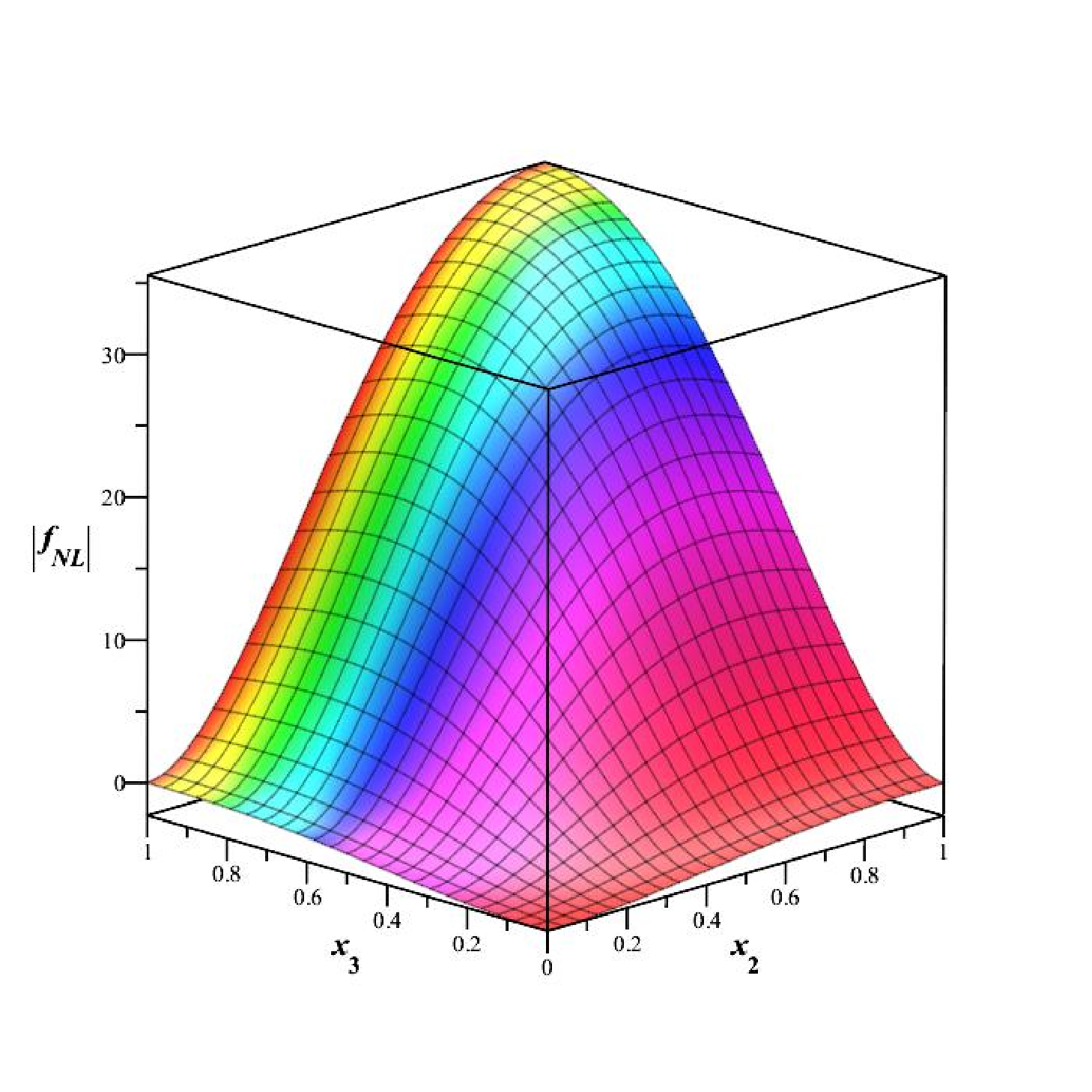}
\caption{\small The non-linearity parameter versus the momenta
ratios $x_{2}$ and $x_{3}$. The plots are corresponding to the cases
with $\beta=0.4$ (up-left panel), $\beta=0.6$ (up-right panel) and
$\beta=0.9$ (lower panel). We see a peak in $x_{2}=x_{3}=1$ for each
panel. } \label{fig9}
\end{figure}

An essential feature of any inflationary scenario, whether isotropic
or anisotropic, is its prediction for the non-Gaussianity of
primordial perturbations. A theoretically consistent model must not
only yield power spectra in agreement with observations, but also
predict an amplitude of non-Gaussianity that lies within current
observational bounds. Motivated by the observationally viable regions obtained from the
scalar and tensor spectra, we next investigate whether the predicted
non-Gaussianity also remains within current observational bounds. To investigate this in detail, we analyze the
behavior of the non-linearity parameter \( f_{NL} \) in the
equilateral configuration, where non-Gaussian features are typically
most pronounced in the DBI-type models. For this purpose, we employ
equation~(\ref{eq44}) and examine how \( f_{NL} \) varies with the
momentum ratios \( x_2 = \frac{k_2}{k_1} \) and \( x_3 =
\frac{k_3}{k_1} \). Figure~\ref{fig9} illustrates this behavior for
three representative values of the intermediate parameter \(\beta\).
As shown in the plots, the amplitude of the non-Gaussianity peaks
around the equilateral limit (\(x_2 = x_3 = 1\)), which justifies focusing on the equilateral configuration in the
subsequent analysis.

\begin{figure}[htbp]
\includegraphics[width=0.33\textwidth]{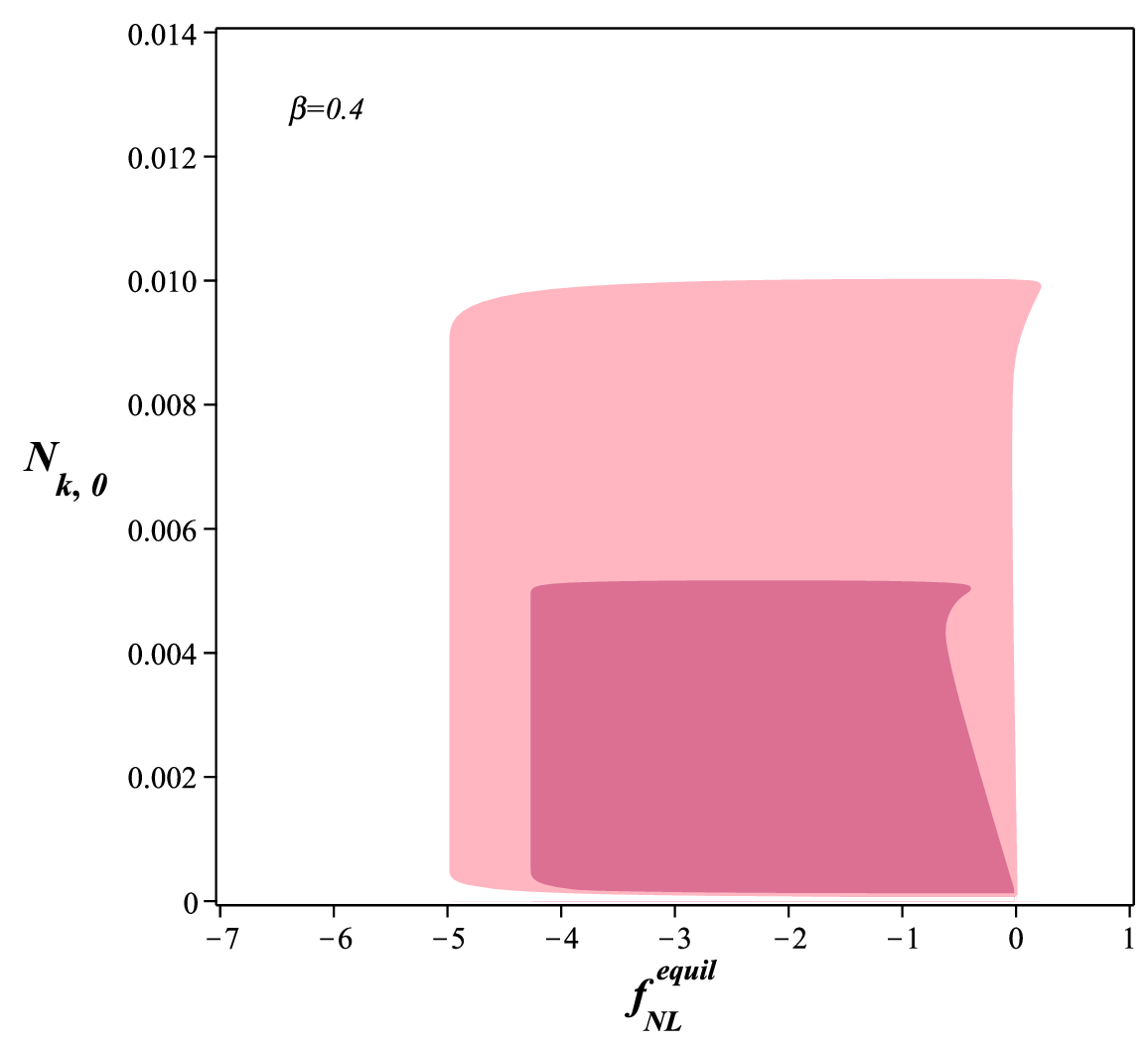} 
\includegraphics[width=0.32\textwidth]{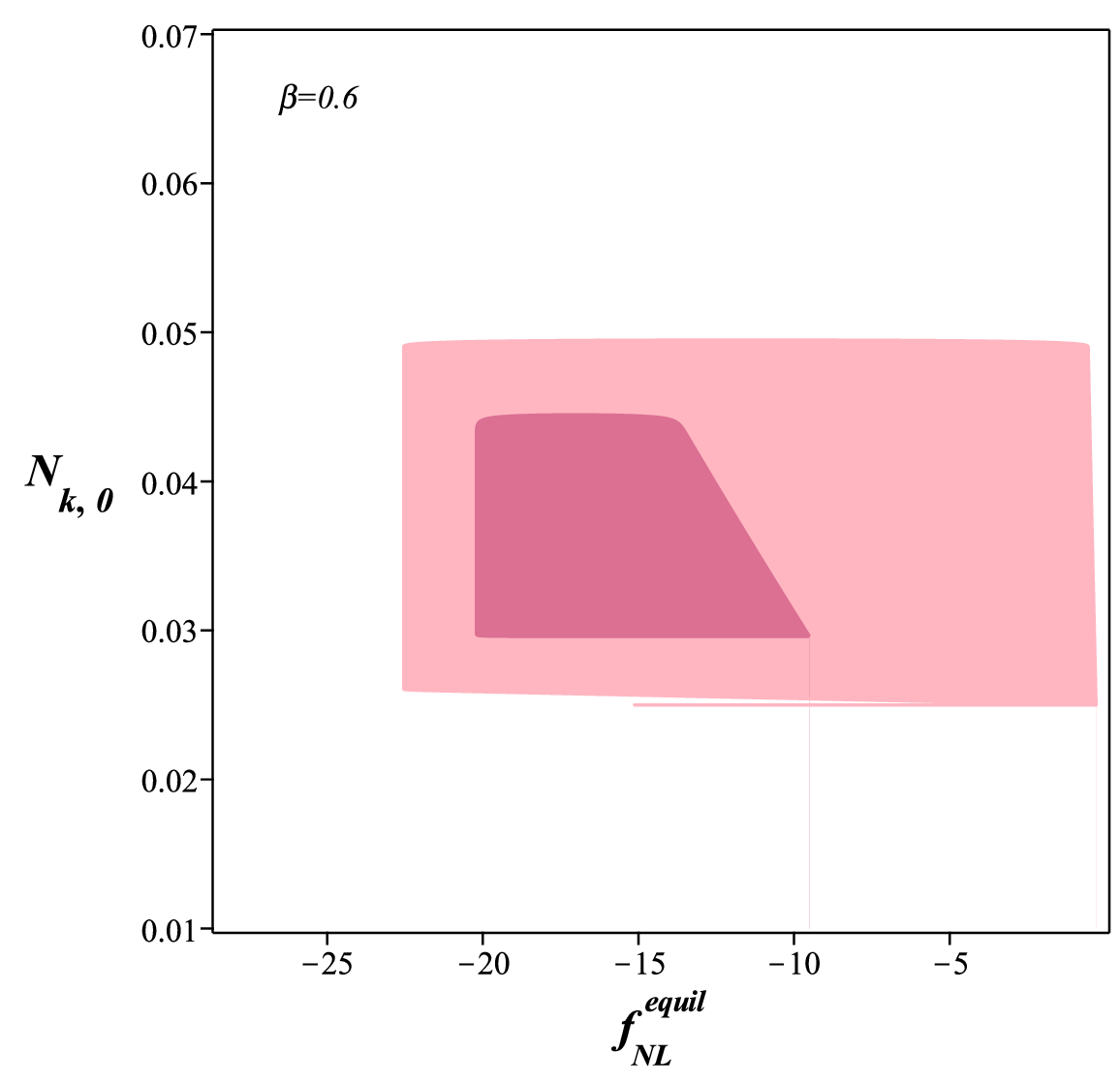} 
\includegraphics[width=0.32\textwidth]{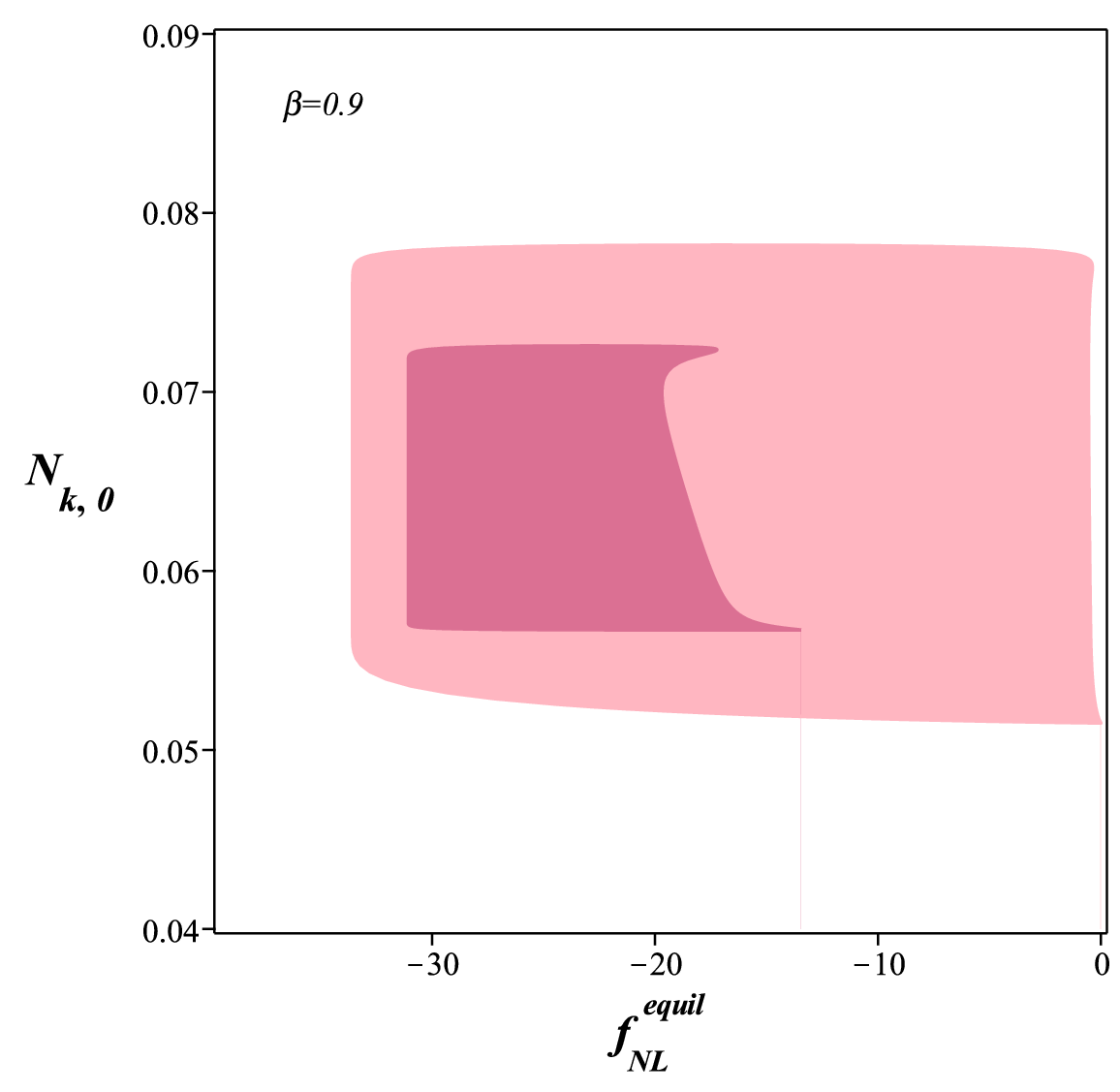}
\caption{\small The parameter space of $N_{k,0}$ and
$f_{NL}^{equil}$,
based on the parameter space allowed by observational limits on the scalar spectral index and the tensor-to-scalar ratio, based on Planck2018 TT,
TE, EE +lowE+lensing+BK18+BAO data.  } \label{fig10}
\end{figure}

\begin{figure}[htbp]
\includegraphics[width=0.33\textwidth]{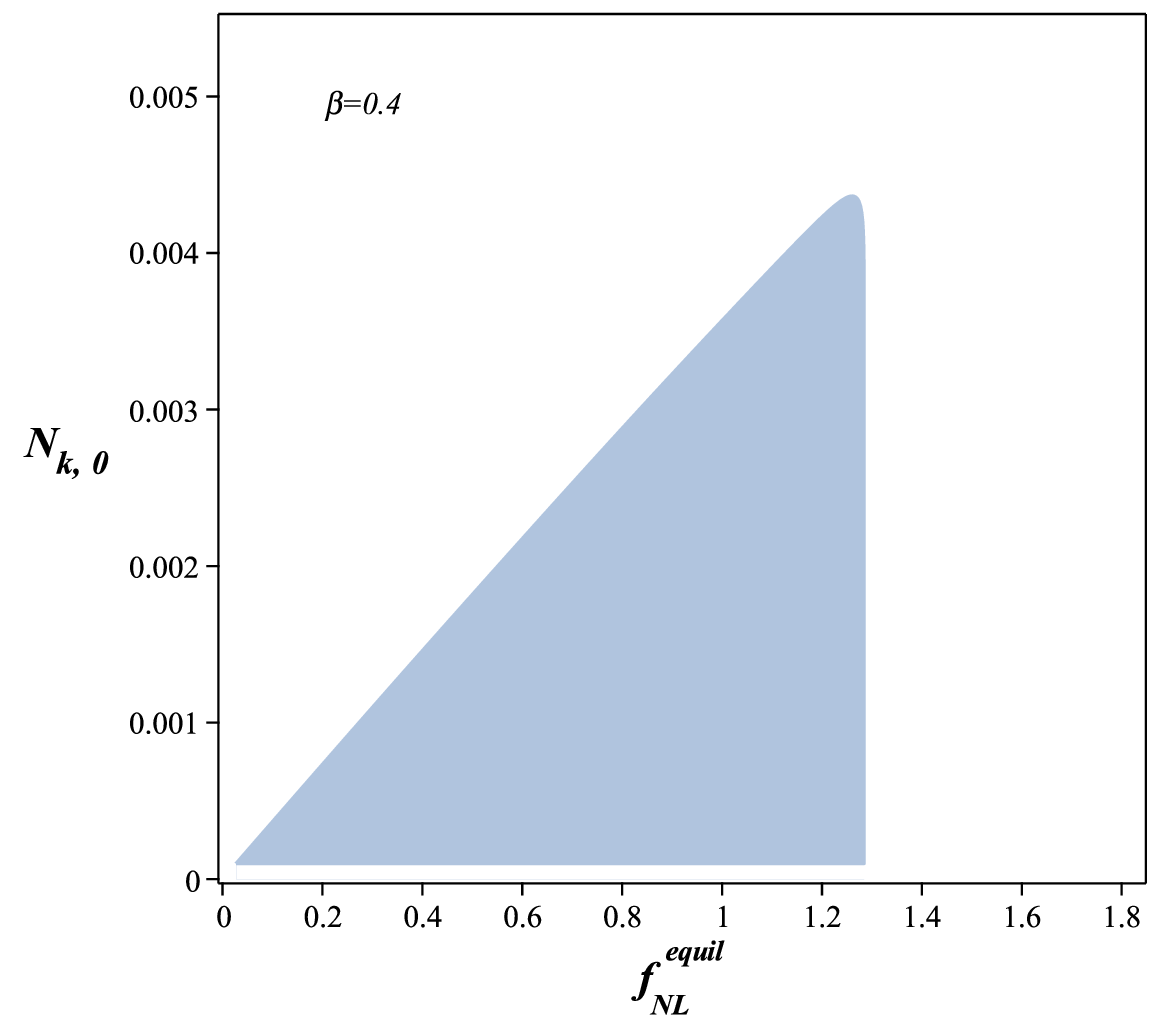} 
\includegraphics[width=0.32\textwidth]{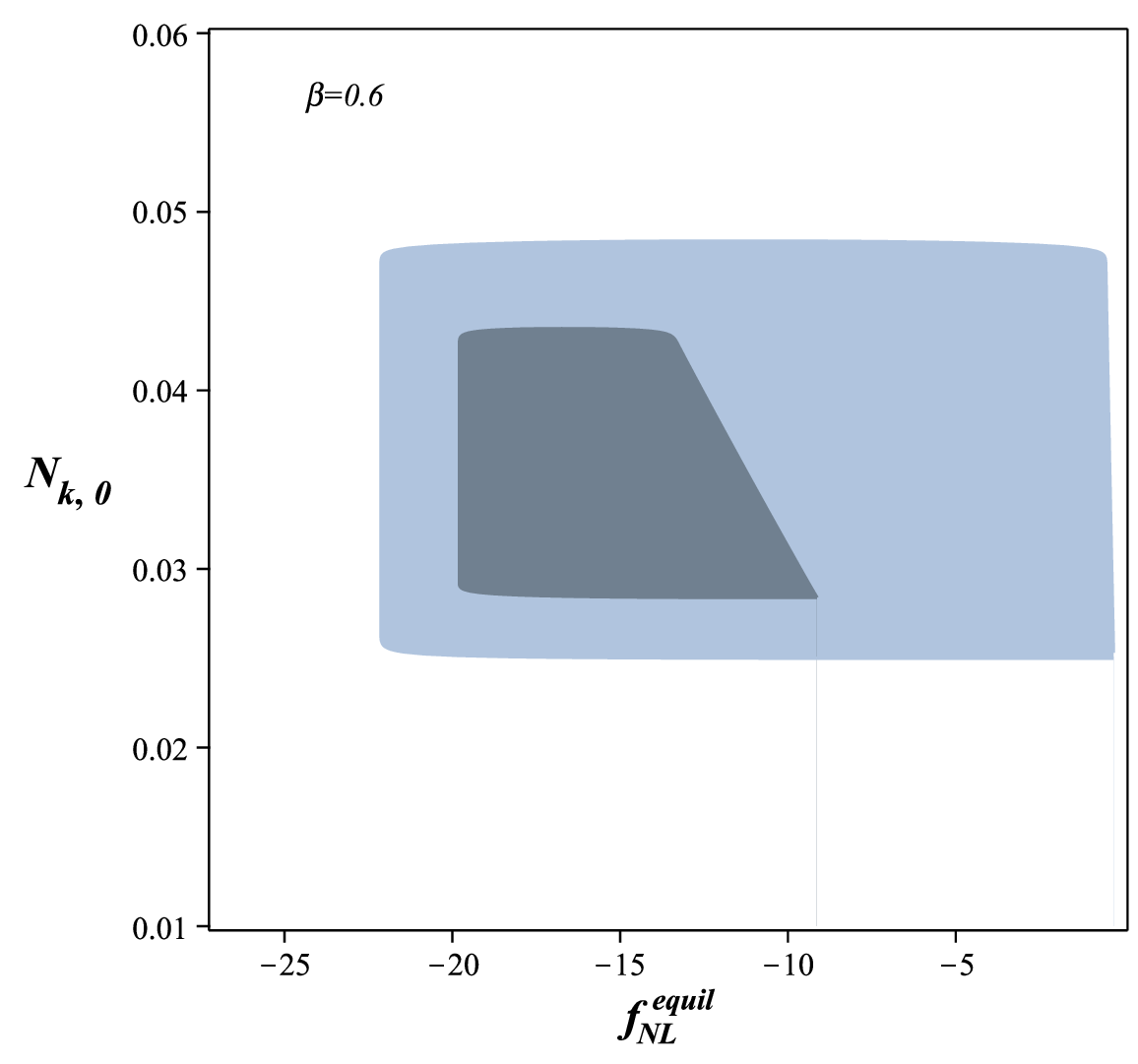} 
\includegraphics[width=0.32\textwidth]{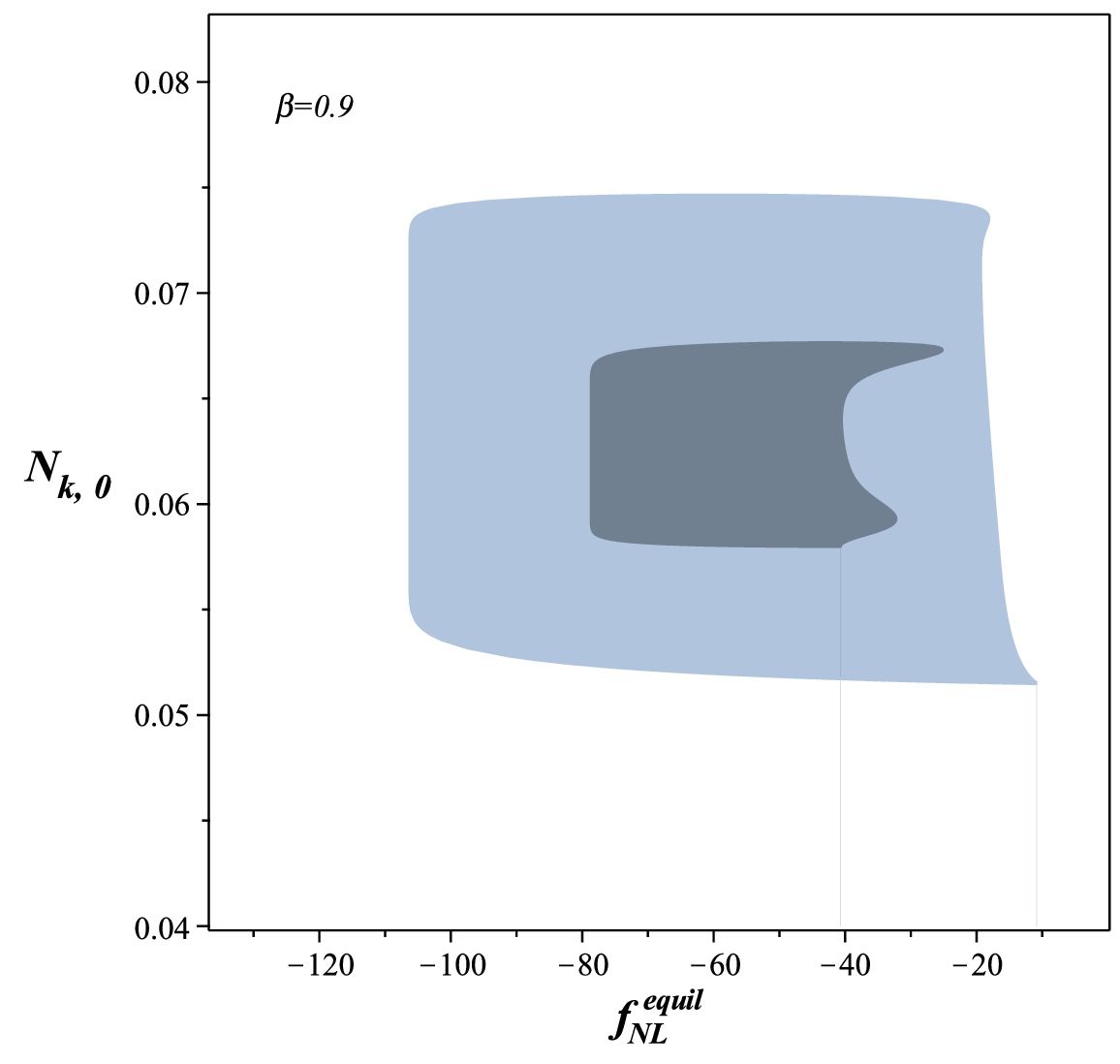}
\caption{\small The parameter space of $N_{k,0}$ and
$f_{NL}^{equil}$, obtained from the observationally viable values of the scalar spectral index and tensor to scalar ratio, based on DESI+CMB+DESY5 data. } \label{fig11}
\end{figure}

\begin{figure}[htbp]
\includegraphics[width=0.32\textwidth]{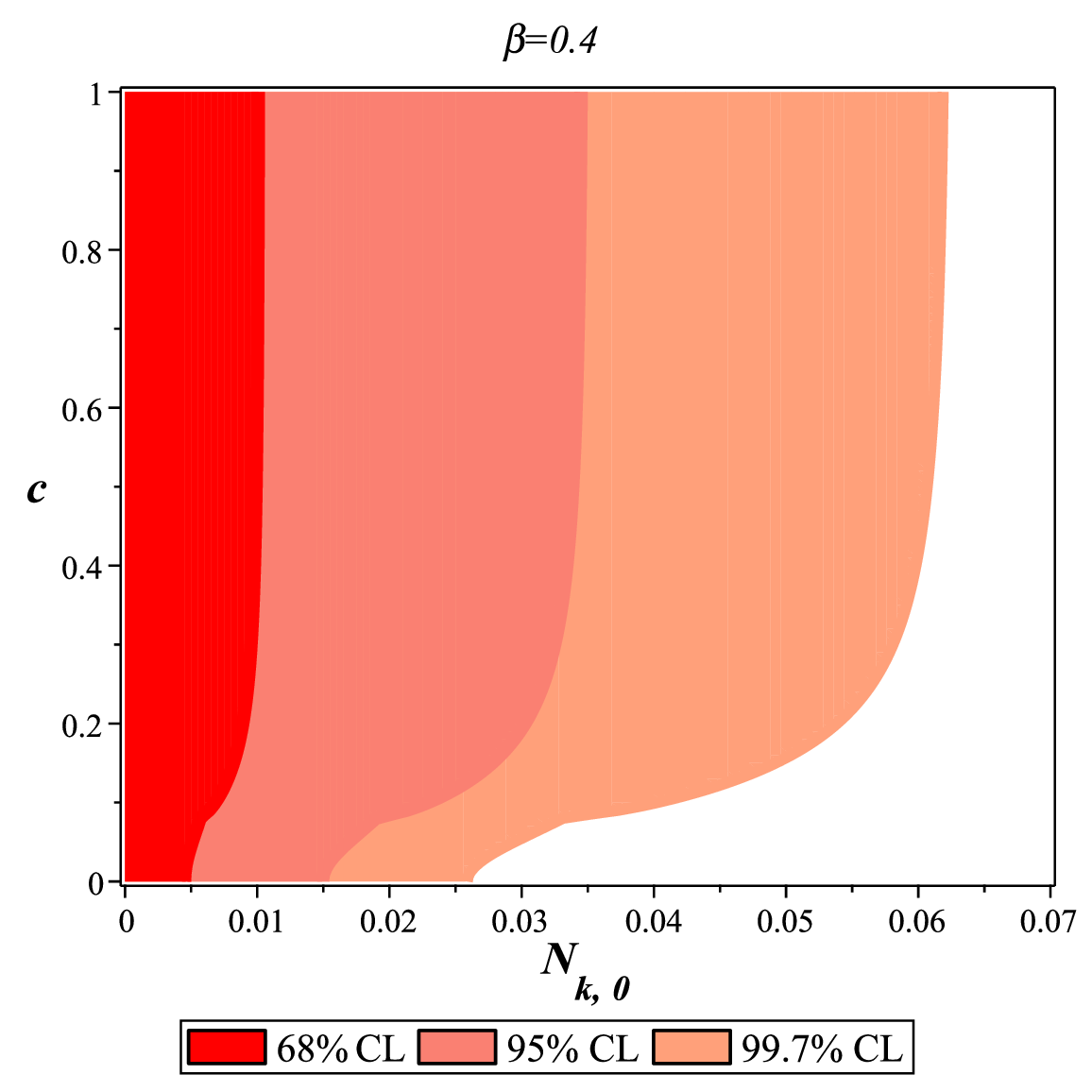} 
\includegraphics[width=0.32\textwidth]{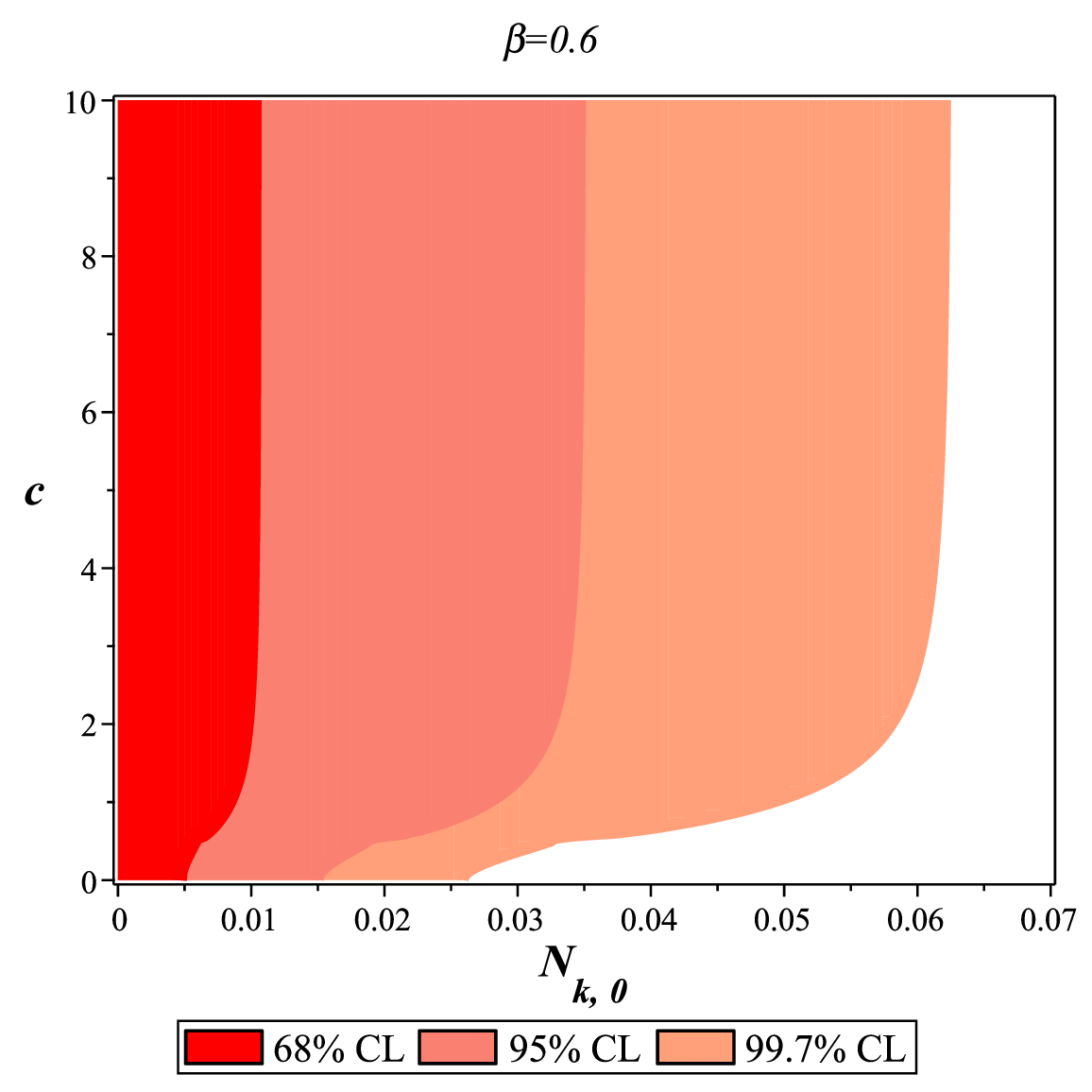} 
\includegraphics[width=0.32\textwidth]{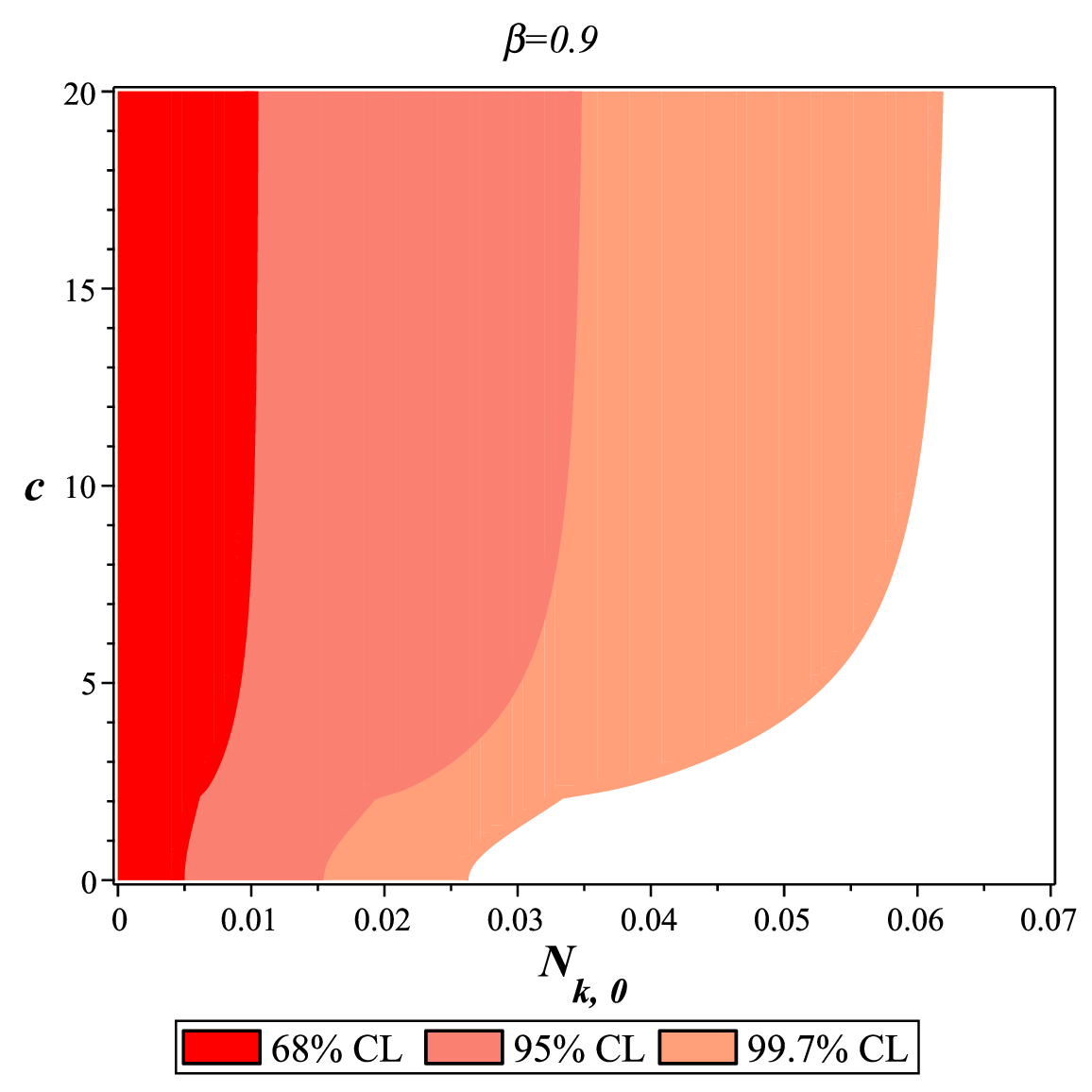}
\caption{\small Allowed regions in the parameter space of $c$ and
$N_{k,0}$, where the predicted equilateral non-Gaussianity amplitude
$f_{NL}^{equil}$ remains consistent with the Planck2018 temperature
and polarization constraints at $68\%$ CL, $95\%$ CL and $99.7\%$
CL. } \label{fig12}
\end{figure}

At this point, we utilize the constraints derived from our numerical
study of the scalar spectral index \(n_s\) and tensor-to-scalar
ratio \(r\) to evaluate the equilateral configuration of the
non-Gaussianity parameter, \(f_{NL}^{\mathrm{equil}}\). Our findings
reveal that the same regions of the parameter space that yield
observationally acceptable values for linear perturbations also lead
to viable predictions for the amplitude of non-Gaussianity.
Figures~\ref{fig10} and~\ref{fig11} illustrate the allowed intervals
of \(f_{NL}^{\mathrm{equil}}\), as inferred from the constraints on
\(r\) and \(n_s\), for both Planck2018 and DESI+CMB+DESY5 datasets.
These results indicate that the model can remain consistent with
current observational bounds on non-Gaussianity. Specifically, the Planck2018
joint temperature and polarization analysis constrains the
equilateral shape of non-Gaussianity to lie within
\(f_{NL}^{\mathrm{equil}} = -26 \pm 47\)~\cite{Pa22}. To provide
more detail, Tables~\ref{tab7}-\ref{tab9} report the allowed values
of the non-linearity parameter for several representative
combinations of model parameters. Moreover, Figure~\ref{fig12}
displays the viable domains in the \(c\)--\(N_{k,0}\) space,
corresponding to the observational limits on
\(f_{NL}^{\mathrm{equil}}\) at confidence levels of 68\%, 95\%, and
99.7\%. These numerical results are further summarized in
Tables~\ref{tab10}-\ref{tab12}. Overall, our analysis suggests that the anisotropic DBI model with
an intermediate expansion scale factor admits regions of parameter
space that are compatible with current observational constraints on
both Gaussian and non-Gaussian observables. Notably, we observe that increasing the
value of the intermediate parameter \(\beta\) leads to a broader
viable range for the model parameter \(c\), while the number density
parameter \(N_{k,0}\) remains within the same order of magnitude
across different confidence levels. At 68\% CL, typical values for
\(N_{k,0}\) are on the order of \(10^{-3}\) to \(10^{-2}\), increasing to
\(\mathcal{O}(10^{-2})\) at 95\% and 99.7\% CL.

\begin{table*}[htbp]
\tiny\tiny\caption{\small{\label{tab7} Viable ranges of the
non-linearity parameter based on different joint data sets. This
table is for $\beta=0.4$}}
\begin{center}
\tabcolsep=0.05cm\begin{tabular}{|c|c|c|c|c|c|} \hline \hline &&&&\\
& Planck2018 TT,TE,EE+lowE & Planck2018 TT,TE,EE+lowE &
DESI+CMB+DESY5 & DESI+CMB+DESY5
\\
& +lensing+BK18+BAO & +lensing+BK18+BAO& &
\\
\hline &&&&\\$N_{k,0}$& $68\%$ CL & $95\%$ CL &$68\%$ CL & $95\%$ CL
\\
\hline\hline &&&&\\  $1.00\times10^{-3}$ &
$-8.509<f_{NL}^{equil}<-0.3271$ & $-4.969<f_{NL}^{equil}<-0.0086
$ & --- & $0.2728<f_{NL}^{equil}<1.284$\\&&&& \\
\hline &&&&\\$4.00\times10^{-3}$& $-8.509<f_{NL}^{equil}<-1.224$ &
$-4.969<f_{NL}^{equil}<-0.0331$ &---& $1.127<f_{NL}^{equil}<1.284$
\\ &&&& \\ \hline &&&&\\
$7.00\times10^{-3}$& --- & $-4.969<f_{NL}^{equil}<-0.0488
$ & --- & --- \\ &&&& \\
\hline &&&&\\
$1.00\times10^{-2}$& --- & $-4.969<f_{NL}^{equil}<0.2377$ & --- &
---
\\ &&&&\\
\hline
\end{tabular}
\end{center}
\end{table*}

\begin{table*}[htbp]
\tiny\tiny\caption{\small{\label{tab8} Viable ranges of the
non-linearity parameter based on different joint data sets. This
table is for $\beta=0.6$}}
\begin{center}
\tabcolsep=0.05cm\begin{tabular}{|c|c|c|c|c|c|} \hline \hline &&&&\\
& Planck2018 TT,TE,EE+lowE & Planck2018 TT,TE,EE+lowE &
DESI+CMB+DESY5 & DESI+CMB+DESY5
\\
& +lensing+BK18+BAO & +lensing+BK18+BAO& &
\\
\hline &&&&\\$N_{k,0}$& $68\%$ CL & $95\%$ CL &$68\%$ CL & $95\%$ CL
\\
\hline\hline &&&&\\  $3.00\times10^{-2}$ &
$-67.33<f_{NL}^{equil}<-9.629$ & $-22.53<f_{NL}^{equil}<-0.3517
$ & $-65.96<f_{NL}^{equil}<-9.627$ & $-22.12<f_{NL}^{equil}<-0.3517$\\&&&& \\
\hline &&&&\\$3.50\times10^{-2}$& $-67.33<f_{NL}^{equil}<-10.01$ &
$-22.53<f_{NL}^{equil}<-0.406$ & $-65.96<f_{NL}^{equil}<-11.10$ &
$-22.12<f_{NL}^{equil}<-0.420$
\\ &&&& \\ \hline &&&&\\
$4.00\times10^{-2}$& $-67.33<f_{NL}^{equil}<-12.56$ &
$-22.53<f_{NL}^{equil}<-0.458 $ & $-65.96<f_{NL}^{equil}<-12.55 $
& $-22.12<f_{NL}^{equil}<-0.4573$ \\ &&&& \\
\hline &&&&\\
$4.50\times10^{-2}$& $-67.33<f_{NL}^{equil}<-13.83 $ &
$-22.53<f_{NL}^{equil}<-0.5053$ &
--- & $-22.12<f_{NL}^{equil}<-0.5045 $
\\ &&&&\\
\hline
\end{tabular}
\end{center}
\end{table*}

\begin{table*}[htbp]
\tiny\tiny\caption{\small{\label{tab9} Viable ranges of the
non-linearity parameter based on different joint data sets. This
table is for $\beta=0.9$}}
\begin{center}
\tabcolsep=0.05cm\begin{tabular}{|c|c|c|c|c|c|} \hline \hline &&&&\\
& Planck2018 TT,TE,EE+lowE & Planck2018 TT,TE,EE+lowE &
DESI+CMB+DESY5 & DESI+CMB+DESY5
\\
& +lensing+BK18+BAO & +lensing+BK18+BAO& &
\\
\hline &&&&\\$N_{k,0}$& $68\%$ CL & $95\%$ CL &$68\%$ CL & $95\%$ CL
\\
\hline\hline &&&&\\  $5.50\times10^{-2}$ & --- &
$-33.56<f_{NL}^{equil}<-0.4345
$ & --- & $-106.1<f_{NL}^{equil}<-15.84$\\&&&& \\
\hline &&&&\\$6.00\times10^{-2}$& $-103.5<f_{NL}^{equil}<-17.50$ &
$-33.56<f_{NL}^{equil}<-0.4973$ & $-196.4<f_{NL}^{equil}<-35.96$ &
$-106.1<f_{NL}^{equil}<-17.27$
\\ &&&& \\ \hline &&&&\\
$6.50\times10^{-2}$& $-103.5<f_{NL}^{equil}<-18.96$ &
$-33.56<f_{NL}^{equil}<-0.5326 $ & $-196.4<f_{NL}^{equil}<-40.63$
& $-106.1<f_{NL}^{equil}<-18.50$ \\ &&&& \\
\hline &&&&\\
$7.00\times10^{-2}$& $-103.5<f_{NL}^{equil}<-19.79 $ &
$-33.56<f_{NL}^{equil}<-0.5600$ &
--- & $-106.1<f_{NL}^{equil}<-19.49 $
\\ &&&&\\
\hline
\end{tabular}
\end{center}
\end{table*}

\begin{table*}[htbp]
\tiny\tiny\caption{\small{\label{tab10}
Model parameter regions consistent with Planck2018 joint temperature and polarization constraints. This table is for $\beta=0.4$}}
\begin{center}
\tabcolsep=0.05cm\begin{tabular}{|c|c|c|c|}  \hline &&&\\$c$& $68\%$
CL & $95\%$ CL &$99.7\%$ CL
\\
\hline\hline &&&\\  $0.1$ & $0<N_{k,0}<0.006$ & $0<N_{k,0}<0.023
$ & $0<N_{k,0}<0.040$ \\&&& \\
\hline &&&\\$0.4$& $0<N_{k,0}<0.009$ & $0<N_{k,0}<0.033$
&$0<N_{k,0}<0.059$
\\ &&& \\ \hline &&&\\
$0.7$& $0<N_{k,0}<0.009$ & $0<N_{k,0}<0.034
$ & $0<N_{k,0}<0.0613$ \\ &&& \\
\hline
\end{tabular}
\end{center}
\end{table*}

\begin{table*}[htbp]
\tiny\tiny\caption{\small{\label{tab11}
Model parameter regions consistent with Planck2018 joint temperature and polarization constraints. This table is for $\beta=0.6$}}
\begin{center}
\tabcolsep=0.05cm\begin{tabular}{|c|c|c|c|}  \hline &&&\\$c$& $68\%$
CL & $95\%$ CL &$99.7\%$ CL
\\
\hline\hline &&&\\  $1$ & $0<N_{k,0}<0.008$ & $0<N_{k,0}<0.028
$ & $0<N_{k,0}<0.049$ \\&&& \\
\hline &&&\\$4$& $0<N_{k,0}<0.010$ & $0<N_{k,0}<0.034$
&$0<N_{k,0}<0.061$
\\ &&& \\ \hline &&&\\
$7$& $0<N_{k,0}<0.010$ & $0<N_{k,0}<0.034
$ & $0<N_{k,0}<0.0617$ \\ &&& \\
\hline
\end{tabular}
\end{center}
\end{table*}

\begin{table*}[htbp]
\tiny\tiny\caption{\small{\label{tab12} Model parameter regions consistent with Planck2018 joint temperature and polarization constraints. This table is for $\beta=0.9$}}
\begin{center}
\tabcolsep=0.05cm\begin{tabular}{|c|c|c|c|}  \hline &&&\\$c$& $68\%$
CL & $95\%$ CL &$99.7\%$ CL\\
\hline\hline &&&\\  $1$ & $0<N_{k,0}<0.004$ & $0<N_{k,0}<0.016
$ & $0<N_{k,0}<0.028$ \\&&& \\
\hline &&&\\$8$& $0<N_{k,0}<0.009$ & $0<N_{k,0}<0.032$
&$0<N_{k,0}<0.057$
\\ &&& \\ \hline &&&\\
$15$& $0<N_{k,0}<0.009$ & $0<N_{k,0}<0.034
$ & $0<N_{k,0}<0.060$ \\ &&& \\
\hline
\end{tabular}
\end{center}
\end{table*}

\newpage
\section{\label{s6}Summary and Conclusion}

In this work, we investigated the dynamics of Dirac-Born-Infeld
(DBI) inflation embedded within an anisotropic background geometry
inspired by Bianchi type I cosmology, together with an
intermediate form for the scale factor. We began by introducing the
DBI action coupled to a spatially anisotropic metric, from which we
derived the corresponding background evolution equations,
including the background Friedmann equations and slow-roll
parameters incorporating anisotropic contributions. Utilizing the
ADM formalism, we decomposed the metric in \(3+1\) form and worked
in the comoving gauge to derive the second-order action for scalar
perturbations. From this, we obtained the equation of motion for
the curvature perturbation \({\cal R}\) in Fourier space. Within
this framework, we also considered deviations from the
Bunch-Davies vacuum by adopting a general excited initial state
parameterized by Bogoliubov coefficients. This enabled us to
compute the scalar and tensor power spectra in the presence of
anisotropic corrections and non-vacuum initial conditions.
Subsequently, we extended our analysis to the nonlinear regime by
expanding the action to third order. Applying the in-in formalism
in the interaction picture, we derived the bispectrum and obtained
a general expression for the non-linearity parameter
\(f_{NL}\) in the equilateral configuration. This expression
incorporates the effects of both anisotropy and non-BD initial
conditions on primordial non-Gaussianity.

For the numerical analysis, we considered an intermediate
expansion scenario for the background evolution together with a
phenomenological ansatz for the number density \(N_k\) of the
excited initial state. Using this setup, we compared the model
predictions with several cosmological datasets, including
Planck2018 and DESI. Through this analysis, we identified
observationally viable regions in the parameter space that are
compatible with current bounds on the scalar spectral index,
tensor-to-scalar ratio, and non-Gaussianity amplitude. Overall,
our results suggest that the intermediate DBI inflation model in
an anisotropic framework with non-standard initial conditions can
remain compatible with current cosmological observations and
provides a useful framework for studying the possible effects of
anisotropy and non-vacuum initial states in the early universe.

Our analysis began by focusing on the inflationary observables,
namely the scalar spectral index \(n_s\) and the tensor-to-scalar
ratio \(r\), which are directly constrained by current
cosmological observations. We studied the behavior of the
\(r\)--\(n_s\) plane in the context of the anisotropic DBI model
with a non-BD initial state by comparing the model
predictions with Planck2018 TT, TE, EE + lowE + lensing + BK18 +
BAO and DESI+CMB+DESY5 datasets. In this analysis, we considered
three representative values of the intermediate parameter,
\(\beta = 0.4\), \(0.6\), and \(0.9\). Our numerical results
indicate that larger values of \(\beta\) generally correspond to
wider observationally viable intervals for the anisotropy
parameter \(c\). For example, based on Planck2018 data at the
95\% confidence level, we obtained
\(0 < c < 0.671\) for \(\beta = 0.4\),
\(0 < c < 6.38\) for \(\beta = 0.6\), and
\(0 < c < 20.4\) for \(\beta = 0.9\).
Similarly, using the DESI+CMB+DESY5 dataset, the corresponding
intervals become
\(0 < c < 3.22\),
\(0 < c < 4.69\), and
\(0 < c < 14.2\),
respectively. Throughout the analysis, the parameter
\(N_{k,0}\), which characterizes the deviation from the
Bunch-Davies vacuum, remained approximately within the range
\(10^{-3} \lesssim N_{k,0} \lesssim 10^{-1}\),
depending on the choice of \(\beta\) and the observational
dataset. We further extended the analysis by incorporating
additional datasets, including SDSS+CMB+Union3,
DESI+CMB+Union3, SDSS+CMB+DESY5, and DESI+CMB+DESY5, in order to
examine the robustness of these results. The same qualitative
trend persists across these datasets: increasing \(\beta\)
typically enlarges the viable range of \(c\), while the allowed
values of \(N_{k,0}\) remain within a comparable order of
magnitude. These results are summarized in the corresponding
figures and tables presenting the observationally viable regions
of parameter space.

We next investigated the non-Gaussian sector of the model.
Since a viable inflationary scenario should produce acceptable
predictions for both the power spectrum and bispectrum, we used
the parameter regions obtained from the linear perturbation
analysis to evaluate the amplitude of primordial
non-Gaussianity. Our numerical analysis shows that the
bispectrum amplitude reaches its maximum near the equilateral
configuration, in agreement with the typical behavior expected
in DBI-type inflationary models. For the quantitative analysis,
we employed the observational bounds derived from
Planck2018 TT, TE, EE + lowE + lensing + BK18 + BAO and
DESI+CMB+DESY5 datasets. According to the Planck2018 combined
temperature and polarization analysis, the equilateral
non-linearity parameter is constrained as
\(f_{NL}^{\mathrm{equil}} = -26 \pm 47\),
and we find that the model predictions remain within this
interval for a substantial region of parameter space.
However, for larger values of the intermediate parameter,
particularly \(\beta = 0.9\), some parameter combinations lead
to \(f_{NL}^{\mathrm{equil}}\) values that exceed the
observationally allowed range when the DESI+CMB+DESY5 dataset is
considered.

To further refine the analysis, we imposed the Planck2018
constraint on \(f_{NL}^{\mathrm{equil}}\) and examined the
corresponding viable domains in the \((c, N_{k,0})\) parameter
space. Our results indicate a consistent overlap between the
regions of parameter space compatible with the observational
bounds on non-Gaussianity and those satisfying the constraints on
the scalar spectral index and tensor-to-scalar ratio. In summary,
the anisotropic DBI inflationary model with a non-BD
initial condition can remain compatible with current cosmological
observations within specific regions of parameter space. The
model is capable of accommodating both power-spectrum
constraints and equilateral-type non-Gaussianity within the
observationally viable parameter ranges, and therefore provides a
useful framework for studying the interplay between anisotropic
inflationary dynamics and non-vacuum initial states.
 \\
 
\textbf{ACKNOWLEDGMENTS}\\
We thank the referee for the very insightful comments that have
improved the quality of the paper considerably.
 \\

\textbf{Data availability}
All relevant analytical expressions and numerical results are included in the manuscript.

\textbf{Code/Software Availability Statement:} No public code repository is associated with this work.

\appendix{\textbf{Appendix}}

\[
\epsilon=
-\frac{
	6\sqrt6\,a_0^3\,\beta\, b
	\left[
	\beta b\,a_0^6(\beta-1)\,\chi^{1+2\beta}
	+\frac{c^2}{2}(a_0^6-1)\,
	\chi^{3+\beta}e^{-6b\chi^\beta}
	\right]
}{
	\chi\left[
	6\beta^2 b^2 a_0^6 \chi^{2\beta}
	+c^2(1-a_0^6)\chi^2 e^{-6b\chi^\beta}
	\right]^{3/2}
}\,\,, \chi \equiv \left(\frac{\mathcal N}{b}\right)^{1/\beta}.
\]

\begin{align}
	\eta
	=&-\frac{12\sqrt6\,a_0^3}{D^{3/2}
		\left[
		c^2 A e^{-6\mathcal N}\mathcal N^{2/\beta}
		+2a_0^6\beta(\beta-1)b^{2/\beta}\mathcal N
		\right]}
	\Bigg\{
	\frac{5}{2}c^2A\beta^2(\beta-1)
	b^{3/\beta}e^{-6\mathcal N}
	\mathcal N^{(2+2\beta)/\beta}
	\nonumber\\
	&+3c^2A a_0^6\beta^3
	b^{3/\beta}e^{-6\mathcal N}
	\mathcal N^{(2+3\beta)/\beta}
	+\frac{1}{3}c^2A\beta(\beta-1)
	\left(\beta-\frac32\right)
	b^{3/\beta}e^{-6\mathcal N}
	\mathcal N^{(2+\beta)/\beta}
	\nonumber\\
	&+\frac14 c^4A^2\beta
	b^{1/\beta}e^{-12\mathcal N}
	\mathcal N^{(4+\beta)/\beta}
	+\left[
	\frac{1}{12}c^4A^2
	b^{1/\beta}e^{-12\mathcal N}
	\mathcal N^{4/\beta}
	+a_0^{12}\beta^4 b^{5/\beta}\mathcal N^3
	\right](\beta-1)
	\Bigg\}.\nonumber
\end{align}
\[
A \equiv a_0^6-1.
\]

\begin{align}
	s
	=&
	\frac{
		36\sqrt6\,c^{2}\beta^{2}
		a_0^{9}e^{9\mathcal N}
		\left(
		FQ+i\sqrt2\,G
		\right)
	}{
		QDRK
	}.
\end{align}
\[
Q\equiv 
\sqrt{
	18\beta^{4}\mathcal N^{4}b^{4/\beta}e^{12\mathcal N}
	-c^{4}\mathcal N^{4/\beta}
},
\]

\[
D\equiv
\left(
6\beta^{2}\mathcal N^{2}b^{2/\beta}e^{6\mathcal N}
+i\sqrt2\,Q
\right)
\left(
\beta^{2}b^{2/\beta}\mathcal N^{2}a_0^{6}e^{6\mathcal N}
-\frac{c^{2}\mathcal N^{2/\beta}}{6}
\right),
\]

\[
R\equiv
\sqrt{
	c^{2}(1-a_0^{6})\mathcal N^{2/\beta}
	+6\beta^{2}b^{2/\beta}\mathcal N^{2}a_0^{6}e^{6\mathcal N}
},
\]

\begin{align}
	F=&
	\frac16 a_0^{6}\beta^{3}c^{2}
	b^{5/\beta}e^{6\mathcal N}
	\mathcal N^{(5\beta+4)/\beta}
	+\beta^{5}b^{7/\beta}e^{12\mathcal N}
	\left(a_0^{6}+\frac12\right)
	\mathcal N^{(7\beta+2)/\beta}
	\nonumber\\
	&+\frac1{18}a_0^{6}\beta^{2}c^{2}(\beta-1)
	b^{5/\beta}e^{6\mathcal N}
	\mathcal N^{(4\beta+4)/\beta}
	-\frac1{36}\beta c^{4}
	b^{3/\beta}
	\left(a_0^{6}+\frac32\right)
	\mathcal N^{(3\beta+6)/\beta}
	\nonumber\\
	&+\frac{\beta-1}{3}
	\Bigg[
	-\frac1{36}c^{4}b^{3/\beta}
	\left(a_0^{6}+\frac32\right)
	\mathcal N^{(2\beta+6)/\beta}
	+\beta^{4}b^{7/\beta}e^{12\mathcal N}
	\left(a_0^{6}+\frac12\right)
	\mathcal N^{(6\beta+2)/\beta}
	\Bigg],
\end{align}

\begin{align}
	G=&
	-\frac32\beta^{7}b^{9/\beta}
	e^{18\mathcal N}
	\mathcal N^{(9\beta+2)/\beta}
	+a_0^{6}\beta^{5}c^{2}b^{7/\beta}
	e^{12\mathcal N}
	\mathcal N^{(7\beta+4)/\beta}
	\nonumber\\
	&+\frac16\beta^{3}c^{4}
	\left(a_0^{3}-\frac12\right)
	\left(a_0^{3}+\frac12\right)
	b^{5/\beta}e^{6\mathcal N}
	\mathcal N^{(5\beta+6)/\beta}
	-\frac1{72}\beta c^{6}
	\left(a_0^{6}+2\right)
	b^{3/\beta}
	\mathcal N^{(3\beta+8)/\beta}
	\nonumber\\
	&+\frac{\beta-1}{3}
	\left[
	-\frac32\beta^{6}b^{9/\beta}
	e^{18\mathcal N}
	\mathcal N^{(8\beta+2)/\beta}
	+c^{2}H
	\right],
\end{align}

\begin{align}
	H=&
	a_0^{6}\beta^{4}b^{7/\beta}
	e^{12\mathcal N}
	\mathcal N^{(6\beta+4)/\beta}
	\nonumber\\
	&-\frac{c^{2}}{72}
	\Big[
	(-12a_0^{6}+3)\beta^{2}
	e^{6\mathcal N}b^{5/\beta}
	\mathcal N^{(4\beta+6)/\beta}
	+b^{3/\beta}c^{2}
	(a_0^{6}+2)
	\mathcal N^{(2\beta+8)/\beta}
	\Big],
\end{align}

\begin{align}
	K=&
	i\sqrt2\,Q
	\Bigg[
	\beta^{2}\mathcal N^{2}
	\left(\frac{3}{4}\right)
		a_0^{6}b^{2/\beta}e^{6\mathcal N}
		+\frac{c^{2}\mathcal N^{2/\beta}}{6}
	\left(
	a_0^{6}+\frac14
	\right)
	\Bigg]
	\nonumber\\
	&+\left(
	-a_0^{6}c^{2}
	+\frac14 c^{2}
	\right)
	\beta^{2}b^{2/\beta}e^{6\mathcal N}
	\mathcal N^{(2+2\beta)/\beta}
-\frac32 a_0^{6}
	\left[
	\beta^{4}\mathcal N^{4}
	b^{4/\beta}e^{12\mathcal N}
	+\frac19 c^{4}
	\mathcal N^{4/\beta}
	\right],
\end{align}


\begin{thebibliography}{100}

\bibitem{Gut81} A. Guth, Phys. Rev. D \textbf{23}, 347 (1981).

\bibitem{Lin82} A. D. Linde, Phys. Lett. B \textbf{108}, 389 (1982).

\bibitem{Alb82} A. Albrecht \& P. J. Steinhardt, Phys. Rev. Lett. \textbf{48}, 1220 (1982).

\bibitem{Lin90} A. D. Linde, Particle Physics and Inflationary Cosmology (Harwood Academic Publishers, Chur, Switzerland) (1990).

\bibitem{Lid00a} A. R. Liddle \& D. Lyth, Cosmological Inflation and Large-Scale Structure, (Cambridge University Press) (2000).

\bibitem{Lid97} J. E. Lidsey, A. R. Liddle, E. W. Kolb, E. J. Copeland, T. Barreiro \& M. Abney, Rev. Mod. Phys. \textbf{69}, 373 (1997).

\bibitem{Muk81} V. F. Mukhanov \& G. V. Chibisov, JETP Lett. \textbf{33}, 532-535 (1981).

\bibitem{Mal03} J. M. Maldacena, JHEP \textbf{0305}, 013 (2003).

\bibitem{Lin83} A. D. Linde, Phys. Lett. B \textbf{129}, 177-181 (1983).

\bibitem{Lin94} A. D. Linde, Phys. Rev. D \textbf{49}, 748-754 (1994).

\bibitem{Arm99} C. Armendariz-Picon, T. Damour \& V. Mukhanov, Phys. Lett. B \textbf{458}, 209-218 (1999).

\bibitem{Sil04} E. Silverstein \& D. Tong, Phys. Rev. D \textbf{70}, 103505 (2004).

\bibitem{Ali04} M. Alishahiha, E. Silverstein \& D. Tong, Phys. Rev. D \textbf{70}, 123505 (2004).

\bibitem{Pl18a} Y. Akrami, M. Ashdown, J. Aumont, C. Baccigalupi \& M. Ballardini, A\&A \textbf{641}, A10 (2020).

\bibitem{Bar13} N. Bartolo, S. Matarrese, M. Peloso \& A. Ricciardone, Phys. Rev. D \textbf{87}, 023504 (2013).

\bibitem{Kan10} S. Kanno, J. Soda \& M. Watanabe, JCAP \textbf{1012}, 024 (2010).

\bibitem{Oh13} J. Ohashi, J. Soda \& S. Tsujikawa, JCAP \textbf{12}, 009 (2013).

\bibitem{La16} S. Lahiri, JCAP \textbf{09}, 025 (2016).

\bibitem{Do21a} T. Q. Do \& W. F. Kao, Phys. Rev. D \textbf{84}, 123009 (2011).

\bibitem{oh13} J. Ohashi, J. Soda \& S. Tsujikawa, 	Phys. Rev. D \textbf{88}, 103517 (2013). 

\bibitem{Ho18} J. Holland, S. Kanno \& I. Zavala, Phys. Rev. D \textbf{97}, 103534 (2018).

\bibitem{Ng21} D. H. Nguyen, T. M. Pham \& T. Q. Do, Eur. Phys. J. C \textbf{81}, 839 (2021).

\bibitem{It18} A. Ito \& J. Soda, Eur. Phys. J. C \textbf{78}, 55 (2018).

\bibitem{Do21} T. Q. Do \& W. F. Kao, Eur. Phys. J. C \textbf{81}, 525 (2021).

\bibitem{Che07} X. Chen, M.-x. Huang, S. Kachru \& G. Shiu, J. Cosmol. Astropart. Phys. \textbf{01}, 002 (2007).

\bibitem{Hol08} R. Holman, A. J. Tolley, JCAP \textbf{0805}, 001 (2008).

\bibitem{Noj22} S. Nojiri, S .D. Odintsov, V. K. Oikonomou \& A. Constantini, Nuclear Physics B \textbf{985}, 116011 (2022).

\bibitem{Pl19} Y. Akrami, F. Arroja, M. Ashdown, J. Aumont \& C. Baccigalupi et al., arXiv:1905.05697 [astro-ph.CO] (2019).

\bibitem{Pa22} D. Paoletti, F. Finelli, J. Valiviita \& M. Hazumi, Phys. Rev. D \textbf{106}, 083528 (2022).

\bibitem{Wa24} D. Wang, arXiv:2404.06796 [astro-ph.CO] (2024).

\bibitem{Co24} S. S. da Costa, arXiv:2412.14290 [astro-ph.CO] (2024).

\bibitem{To04} D. Tong, in \textit{Proceedings of the 12th International Conference on Supersymmetry and Unification of Fundamental Interactions (SUSY 2004)}, Tsukuba, Japan, 17--23 June 2004, pp. 841--844.

\bibitem{Li13} S. Li \& A. R. Liddle, JCAP \textbf{03}, 044 (2014).

\bibitem{Ch08} L. P. Chimento \& R. Lazkoz, Gen. Rel. Grav. \textbf{40}, 2543-2555 (2008).

\bibitem{Ad25} A. G. Adame, J. Aguilar, S. Ahlen, S. Alam \& D. M. Alexander et al., JCAP \textbf{02}, 021 (2025).

\bibitem{Ad24} A. G. Adame, J. Aguilar, S. Ahlen, S. Alam, D. M. Alexander et al., arXiv:2411.12022 [astro-ph.CO] (2024).


\bibitem{Fel11} A. De Felice \& S. Tsujikawa, JCAP \textbf{1104}, 029 (2011).

\bibitem{Gan11} J. Ganc, Phys. Rev. D \textbf{84}, 063514 (2011).





\end{thebibliography}
\end{document}